\documentclass[12pt]{Utils/wlscirep}
\usepackage[utf8]{inputenc}
\usepackage{float}
\usepackage[T1]{fontenc}
\usepackage[draft]{changes}

\usepackage{Utils/math_commands}
\usepackage{graphicx}%
\usepackage{multirow}%
\usepackage{mathrsfs}%

\usepackage{algorithm}%
\usepackage{algorithmicx}%
\usepackage{algpseudocode}%

\usepackage[title]{appendix}%
\usepackage{xcolor}%
\usepackage{textcomp}%
\usepackage{manyfoot}%
\usepackage{booktabs}%
\usepackage{listings}%
\usepackage{lineno}
\usepackage[square,comma,numbers,sort]{natbib}

\usepackage{listings}
\definecolor{kleinblue}{rgb}{0,0.18,0.65}
\hypersetup{
colorlinks=true,
linkcolor=magenta,
citecolor=blue,
urlcolor=kleinblue
}
\newcommand{\UpperRomanNumeral}[1]{\uppercase\expandafter{\romannumeral#1}}
\newcommand{\Rst}{\UpperRomanNumeral{1}}
\newcommand{\Rnd}{\UpperRomanNumeral{2}}

\lstdefinestyle{mystyle}{
    backgroundcolor=\color{backcolour},   
    commentstyle=\color{codegreen},
    numberstyle=\tiny\color{codegray},
    stringstyle=\color{codepurple},
    basicstyle=\ttfamily\footnotesize,
    breakatwhitespace=false,         
    breaklines=true,                 
    captionpos=b,                    
    keepspaces=true,                 
    showspaces=false,                
    showstringspaces=false,
    showtabs=false,                  
    tabsize=2
}

\usepackage{hyperref}

\usepackage{zref-user}
\usepackage{xurl}

\usepackage{bibunits}
\usepackage{thm-restate}
\usepackage{multirow}
\usepackage{colortbl}
\usepackage{titletoc} %
\usepackage{etoolbox}
\usepackage{graphicx}
\usepackage{arydshln}
\usepackage{makecell}
\usepackage{tocloft}
\usepackage{lineno}
\usepackage{marvosym}

\makeatletter

\newcommand{\settitle}{\@maketitle} %

\renewcommand{\keywordname}{} 

\makeatother

\makeatletter
\patchcmd{\listof}%
  {\float@listhead}%
  {\@namedef{l@#1}{\l@supfigure}\float@listhead}%
  {}{}%
\makeatother

\begin{document}

\title{Siamese Foundation Models for Crystal Structure Prediction}

\author[1,2,3 \dag]{Liming Wu}
\author[1,2,3
\dag, \Letter]{Wenbing Huang}
\author[4,5]{Rui Jiao}
\author[6]{Jianxing Huang}
\author[6]{Liwei Liu}
\author[6]{Yipeng Zhou}
\author[1,2,3]{Hao Sun}
\author[4,5]{Yang Liu}
\author[4]{Fuchun Sun}
\author[7,
\Letter]{Yuxiang Ren}
\author[1,2,3,
\Letter]{Ji-Rong Wen}
\affil[1]{Gaoling School of Artificial Intelligence, Renmin University of China, Beijing, China}
\affil[2]{Beijing Key Laboratory of Research on Large Models and Intelligent Governance}
\affil[3]{Engineering Research Center of Next-Generation Intelligent Search and Recommendation, MOE}
\affil[4]{Department of Computer Science and Technology, Tsinghua University, Beijing, China}
\affil[5]{Institute for AI Industry Research, Tsinghua University, Beijing, China}
\affil[6]{Advanced Computing and Storage Lab, Huawei Technologies, Shanghai, China}
\affil[7]{School of Intelligence Science and Technology, Nanjing University, Suzhou, China}

\affil[
\dag]{These authors contributed equally: Liming Wu, Wenbing Huang}
\affil[
\Letter]{Correspondence should be addressed to: hwenbing@ruc.edu.cn; renyuxiang@nju.edu.cn; jrwen@ruc.edu.cn}

\begin{abstract}
Predicting crystal structures from chemical compositions is a fundamental challenge in materials discovery, complicated by complex 3D geometries that distinguish it from fields like protein folding. Here, we present Diffusion-based Crystal Omni (DAO), a pretrain–finetune framework for crystal structure prediction integrating two Siamese foundation models: a structure generator and an energy predictor. The generator is pretrained via a two-stage pipeline on a vast dataset of stable and unstable structures, leveraging the predictor to relax unstable configurations and guide the generative sampling. Across two  well-known benchmarks, pretraining significantly enhances performance across multiple backbone architectures. Ablation studies confirm that the synergy between the generator and predictor mutually benefits both components. We further validate DAO on three real-world superconductors ($\text{Cr}_6\text{Os}_2$, $\text{Zr}_{16}\text{Rh}_8\text{O}_4$, and $\text{Zr}_{16}\text{Pd}_8\text{O}_4$) typically inaccessible to conventional computation. For $\text{Cr}_6\text{Os}_2$, DAO achieves a 100\% match rate with experimental references and an atomic-position error of 0.0012 under 20-shot generation, performing over 2000$\times$ faster per iteration than DFT-based structure predictors. These compelling results collectively highlight the potential of our approach for advancing materials science research.

\end{abstract}

\settitle

\begin{bibunit}[unsrt]


\section{Introduction} \label{sec:intro}

Crystals are solid materials composed of atoms, molecules, or ions arranged in an ordered lattice that repeats periodically in three-dimensional space. This highly symmetric, repeating atomic structure endows crystals with unique physical and chemical properties, making them indispensable in advanced technologies and applications, such as the design of superconductors, ferroelectrics and catalysts~\cite{buckel2008superconductivity,cochran1960crystal,larsen1999fundamental}. Therefore, Crystal Structure Prediction (CSP), which determines the stable 3D structure of a compound solely from its chemical composition, has remained a fundamental and long-standing pursuit since its conceptual inception in the 1950s~\cite{desiraju2002cryptic}. The significance of CSP in materials science is analogous to the well-established field of protein structure prediction (or protein folding) in biology. For protein structure prediction, remarkable advancements have been achieved through tools like the AlphaFold series~\cite{senior2020alphafold,jumper2021alphafold2,abramson2024alphafold3}; in contrast, CSP remains a relatively underexplored task owing to the much more complicated geometries of crystal structures.

Traditional methods for CSP, such as first-principles calculations~\cite{kohn1965dft}, stochastic sampling~\cite{pannetier1990prediction}, and evolutionary optimization~\cite{wang2012calypso}, generally employ either physics-based or data-driven strategies. These methods, while valuable, are inherently limited by
high computational costs and poor scalability with system  complexity. In light of the inherent limitations of traditional CSP methods, deep learning has emerged as a powerful technique for achieving a more favorable balance between accuracy and computational cost. Specifically, deep generative models, such as diffusion and flow models~\cite{ho2020ddpm,song2021maximum,lipman2022flow}, have been employed to learn the underlying distribution of crystal structures from existing databases~\cite{nouira2018crystalgan,xie2022cdvae,jiao2024diffcsp,jiao2024diffcsp++,luo2024crystalflow,liu2024shotgun,wu2026dmflow}. A key advantage of deep generative models, as highlighted in~\cite{jiao2024diffcsp}, lies in the denoising process which functions similarly to a force field by guiding atom coordinates toward local energy minima, thereby enhancing structural stability. In addition, active learning~\cite{merchant2023GNoME,hessmann2025accelerating,butler2024machine} enables exploration of uncharted spaces without database dependence. Nevertheless, these State-Of-The-Art (SOTA) models still fall short of achieving satisfactory performance on widely recognized CSP benchmarks like MPTS-52~\cite{jain2013_mp_dataset}, majorly due to their reliance on domain-specific small datasets for training and the limited generalizability to unseen structures. 

One promising approach to enhancing the generalizability of CSP models could involve leveraging foundation models~\cite{bommasani2021foundation_model}. These models, which are pretrained on extensive datasets and finetuned for specific domains, have shown the great power of emergence and homogenization, establishing themselves as a central paradigm in modern AI systems~\cite{liu2024deepseekv3,achiam2023_gpt4}. Inspired by this trend, the development of foundation models for crystals is gaining prominence as a critical research direction in materials science. Existing crystal foundation models can be broadly classified into two categories: supervised and self-supervised pretraining approaches. The first category pretrains models on crystals with energy and force labels, aiming at learning inter-atomic potentials. Representative examples include GNoME~\cite{merchant2023GNoME}, DPA-2~\cite{zhang2023dpa}, MatterSim~\cite{yang2024mattersim}, and MACE-MP-0~\cite{batatia2023macp_mp_0}. Despite their achievements, these methods are primarily applied to predict the force field of crystals, which deviates from the CSP objective of identifying stable structures. The second category adopts self-supervised pretraining through either 
predictive objectives (e.g., feature reconstruction~\cite{magar2022crystaltwins,yu2023mmpt,das2023crysgnn} and coordinate denoising~\cite{koker2022CrystalCLR,new2024self}) or generative modeling~\cite{zeni2023mattergen}. However, none of these methods specifically targets CSP as a downstream application. Predictive approaches are designed for crystal property prediction, while the generative method MatterGen~\cite{zeni2023mattergen} focuses on general-purpose crystal generation under conditions of desired chemistry, symmetry, and properties. Although MatterGen can be adapted for CSP under the condition of desired chemistry and low-energy, this capability remains peripheral to the study's central objectives and is not thoroughly investigated. 

In this paper, we propose Siamese foundation models specifically designed to tackle CSP. Our proposed framework termed Diffusion-based crystAl Omni (DAO), comprises two complementary foundation models: DAO-G, which is responsible for Generating stable structures; and DAO-P, which specializes in Predicting energy and assisting DAO-G. In particular, DAO-G is a generative model that directly employs CSP as its pretraining task using the diffusion process from DiffCSP~\cite{jiao2024diffcsp}. To enable DAO-G to learn from a broader distribution, we incorporate both stable and unstable structures during pretraining. To facilitate this, we first compile a large pretraining dataset, CrysDB, comprising approximately 940K entries of stable and unstable crystals with energy annotations. Using CrysDB, we pretrain DAO-G through a two-stage pipeline: in Stage \Rst, DAO-G is trained on all crystals in CrysDB, and in Stage \Rnd, it is further trained on a relaxed dataset, where DAO-P acts as an energy predictor to refine unstable structures into more stable conformations. During the generation process of DAO-G, DAO-P further steers the generated structures via energy guidance. To equip DAO-P with these capabilities, we pretrain it on CrysDB through supervised energy prediction and self-supervised structure generation based on DiffCSP~\cite{jiao2024diffcsp}. Notably, both DAO-P and DAO-G are built upon our proposed geometric graph Transformer, \emph{i.e.} Crysformer, which ensures the necessary $\mathrm{O}(3)$ and periodic invariance for crystal structures. 

We finetune and evaluate our models on two well-known CSP benchmarks, the MP-20 and MPTS-52 datasets~\cite{jain2013_mp_dataset}. The results demonstrate that pretraining consistently improves performance for multiple backbone architectures, validating the effectiveness of our proposed methodology. Extensive studies further confirm that DAO-G excels in generating diverse polymorphic structures, and the dataset relaxation and energy guidance provided by DAO-P are essential for enhancing DAO-G's performance. As an add-on benefit, our pretrained DAO-P can also be applied to downstream crystal property prediction tasks, achieving SOTA results on four datasets and ranking within the top three on three others. Finally, we apply our models to superconducting materials. The pursuit of high-temperature superconductors is motivated by their potential applications in efficient energy transmission and quantum computing~\cite{ladd2010quantum}. However, the complex structures of superconducting materials have long made their design a formidable challenge. To address this, we finetune DAO-G on a 3D superconductor dataset curated from the SuperCon source~\cite{supercon} and subsequently finetune DAO-P to estimate the critical temperature $T_c$ using augmented structures generated by DAO-G. Remarkably, for three real-world superconductors not included in the pretraining and finetuning processes, DAO-G accurately and efficiently predicts their structures. For instance, on $\text{Cr}_6\text{Os}_2$~\cite{flukiger1974electronically}, DAO-G generates structures very close to the experimental one with a 100\% Match Rate and an RMSE of 0.0012 over 20 runs. Moreover, we employ Density Functional Theory (DFT)~\cite{hafner2008ab} to compute the energy above hull (Ehull) for both the generated and ground-truth structures. The resulting error is no more than $2\times10^{-5}$ eV/atom, confirming the thermodynamic stability of generated structures. On the contrary, conventional DFT-based structure predictors derive the structure of $\text{Cr}_6\text{Os}_2$, taking more than 2000 times longer per iteration. For $\text{Zr}_{16}\text{Rh}_8\text{O}_4$ and $\text{Zr}_{16}\text{Pd}_8\text{O}_4$, DAO-P predicts their critical temperatures with high accuracy, achieving errors of 0.26 K and 0.04 K, respectively. These compelling results collectively highlight the potential of our approach for advancing materials science research and development.

In summary, this work introduces the Siamese foundation models, DAO-G and DAO-P, designed to tackle the CSP problem. DAO-G generates stable structures, while DAO-P assists by predicting energy and guiding the generative process. Pretraining on a large dataset, CrysDB, enables DAO-G to learn from both stable and unstable crystal structures. Extensive evaluation on CSP benchmarks demonstrates the effectiveness of DAO-G, especially in generating diverse polymorphic structures, with DAO-P's energy guidance significantly enhancing performance. Additionally, DAO-P achieves SOTA results in crystal property prediction. Applied to superconducting materials, DAO-G accurately predicts structures and critical temperatures for real-world superconductors, outperforming conventional methods in both efficiency and accuracy.

\begin{figure}[t!]
\centering
\includegraphics[scale=0.86]{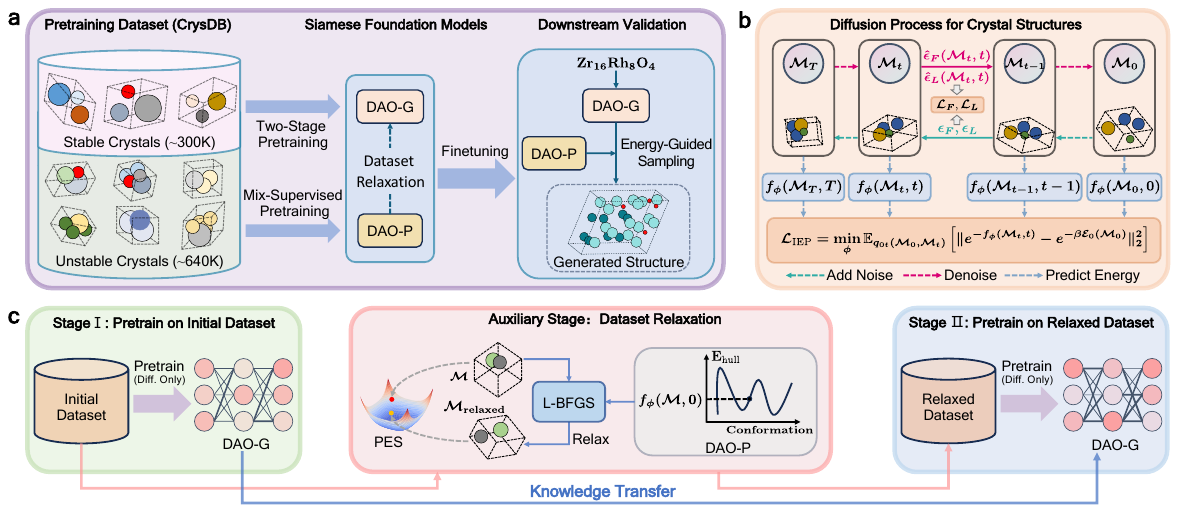}
\caption{A summary of our DAO. \textbf{a}, The pretraining process and downstream validation of our two complementary foundation models: DAO-G and DAO-P. \textbf{b}, The mix-supervised pretraining for DAO-P. It involves a diffusion-based CSP loss (self-supervised) to estimate the lattice noise and fractional coordinates score, and an exponential energy loss (supervised) aiming at recovering the intermediate energy at each timestep along the diffusion trajectory. \textbf{c}, The two-stage pretraining pipeline for DAO-G.
In Stage \Rst, DAO-G is pretrained using a diffusion-based CSP loss on the original dataset. Then, DAO-P is employed to relax unstable structures. In Stage \Rnd, DAO-G is continually pretrained on the relaxed dataset.}
\label{fig:model}
\end{figure}

\section{Results}\label{sec:results}
\subsection{The Pretrain-Finetune Framework of Our Models}\label{sec:pretrain_finetune_framework}
A crystal is represented as $\gM=(\rmA, \vec{\rmL}, \rmF)$, where $\vec{\rmL}\in\sR^{3\times 3}$ is the lattice, $\rmA\in\sR^{1\times N}$ and $\rmF\in[0,1)^{3\times N}$ denote the types and fractional coordinates of $N$ atoms within the lattice, respectively. A crystal $\gM$ is typically considered stable if its energy above the convex hull, namely Ehull, is no greater than 0.08 eV/atom. The CSP task is formulated as learning the distribution $p(\vec{\rmL}, \rmF|\rmA)$, which generates the stable structure $(\vec{\rmL},\rmF)$ of a crystal given its chemical composition $\rmA$. To tackle CSP, we design Siamese foundation models: DAO-G for structure generation and DAO-P for energy prediction (\cref{fig:model}a). Both models are built upon a Transformer-like architecture called Crysformer (\cref{fig:crysformer}), which comprises an embedding module, an invariant graph attention module, a gated addition module, and noise/energy output heads. Through its careful design, Crysformer effectively characterizes the geometry of the input crystal while capturing crystallographic symmetries---ensuring that the noise output is equivariant and the energy output is invariant to arbitrary E(3) transformations and periodic translations of the input structure. Further details are provided in \cref{sec:model_invariance_proof,sec:noise_equivariance_proof}. 
We next describe the pretrain-finetune details of DAO-G and DAO-P, separately.

We conduct a two-stage pretraining pipeline for DAO-G (\cref{fig:model}c). In Stage \Rst, in order to learn from a wider distribution, DAO-G is trained to address CSP on all crystals in CrysDB (introduced in~\cref{sec:pretraining_dataset}), which contains a considerable proportion of unstable structures. The CSP task is implemented via a diffusion process proposed by DiffCSP~\cite{jiao2024diffcsp}, where the noise head of DAO-G is required to estimate the lattice noise $\epsilon_L(\gM_t, t)$ and the fractional coordinates score $\epsilon_F(\gM_t, t)$ of the crystal $\gM_t$ at timestep $t$ in the denoising process. The inclusion of unstable structures enables DAO-G to learn from a broader dataset; however, it also introduces a potential bias to unstable regions of the energy landscape, which may limit its effectiveness in generating stable structures. 
To mitigate this issue, we utilize DAO-P as an efficient energy predictor in place of traditional expensive DFT-based calculators~\cite{kohn1965dft}, and relax unstable structures with energy values within the range (0.08, 0.5] eV/atom towards more stable configurations. Specifically, DAO-P calculates their energy gradients (i.e., force fields), based on which structure relaxation is subsequently performed using the L-BFGS optimizer~\cite{liu1989lbfgs}. Then, in Stage \Rnd, we complete DAO-G's pretraining by refining the model from Stage \Rst\space on the relaxed dataset. After training, DAO-G can sample structures by evolving from a prior distribution at $t=T$ to the data distribution at $t=0$. During this sampling process, we further utilize DAO-P as an energy guider to steer the structures generated by DAO-G toward the equilibrium distribution.
More details are provided in \cref{sec:two_stage_pretraining}.

Although DAO-G is also a generative model similar to MatterGen~\cite{zeni2023mattergen}, it diverges in two critical aspects. First, unlike MatterGen which considers both crystal composition and structure generation as pretraining objectives, DAO-G directly employs CSP as its pretraining task using DiffCSP~\cite{jiao2024diffcsp}. Second, while MatterGen pretrains exclusively on stable crystals, our approach incorporates both stable and unstable structures, allowing the models to learn from a broader distribution. 

As mentioned above, DAO-P serves as the energy predictor, playing a dual role in relaxing the pretraining data and guiding the sampling process for DAO-G. To enable this capability, we pretrain DAO-P on CrysDB with two types of loss (\cref{fig:model}b): (1) the CSP loss applied to the noise head, based on a diffusion process similar to DAO-G, and (2) the energy prediction loss applied to the energy head. The second loss, contributing to energy guidance during the sampling process of DAO-G, requires DAO-P to predict the energy of perturbed crystals at each timestep $t$ along the diffusion trajectory, namely the intermediate energy $\gE_t(\gM_t,t)$. However, estimating the intermediate energy is more challenging than predicting the energy of equilibrium structures, as  ground-truth values for intermediate states are not readily available. To overcome this limitation, we propose an exponential energy loss, whose optimal solution is theoretically proved to converge to the real ground-truth energies under the Boltzmann-constrained modeling~\cite{lu2023cep}. More details are deferred to \cref{sec:daop_pretraining}.

After pretraining, DAO-G is well-suited for the CSP task and can be readily finetuned without modifying its architecture. This seamless transition stems from the close alignment between the pretraining process and the CSP objective. We first introduce the CrysDB pretraining dataset (\cref{sec:pretraining_dataset}) and quantify the performance gains conferred by Stage \Rst's pretraining on a deduplicated version of CrysDB for both our framework and baseline methods (\cref{sec:exp_structure_generation}). We then investigate the advantages of Stage \Rnd's pretraining that incorporates unstable structures relaxed via DAO-P (\cref{sec:exp_relaxation}). Furthermore, we demonstrate DAO-G’s capacity for generating polymorphic structures (\cref{sec:exp_poly}) and illustrate how energy-based guidance further prioritizes the generation of thermodynamically stable materials (\cref{sec:exp_guidance}). Beyond structural generation, we assess the versatility of DAO-P in predicting diverse material properties by finetuning it across eight distinct datasets, each utilizing a specialized prediction head (\cref{sec:exp_property_prediction}). Crucially, DAO-G and DAO-P can synergistically collaborate to facilitate the analysis of superconductors (\cref{sec:exp_supercon}): DAO-G is finetuned to generate 3D superconducting architectures, while DAO-P leverages these augmented structures to accurately estimate critical temperatures ($T_c$).

\begin{figure}[t!]
\centering
\includegraphics[scale=0.85]{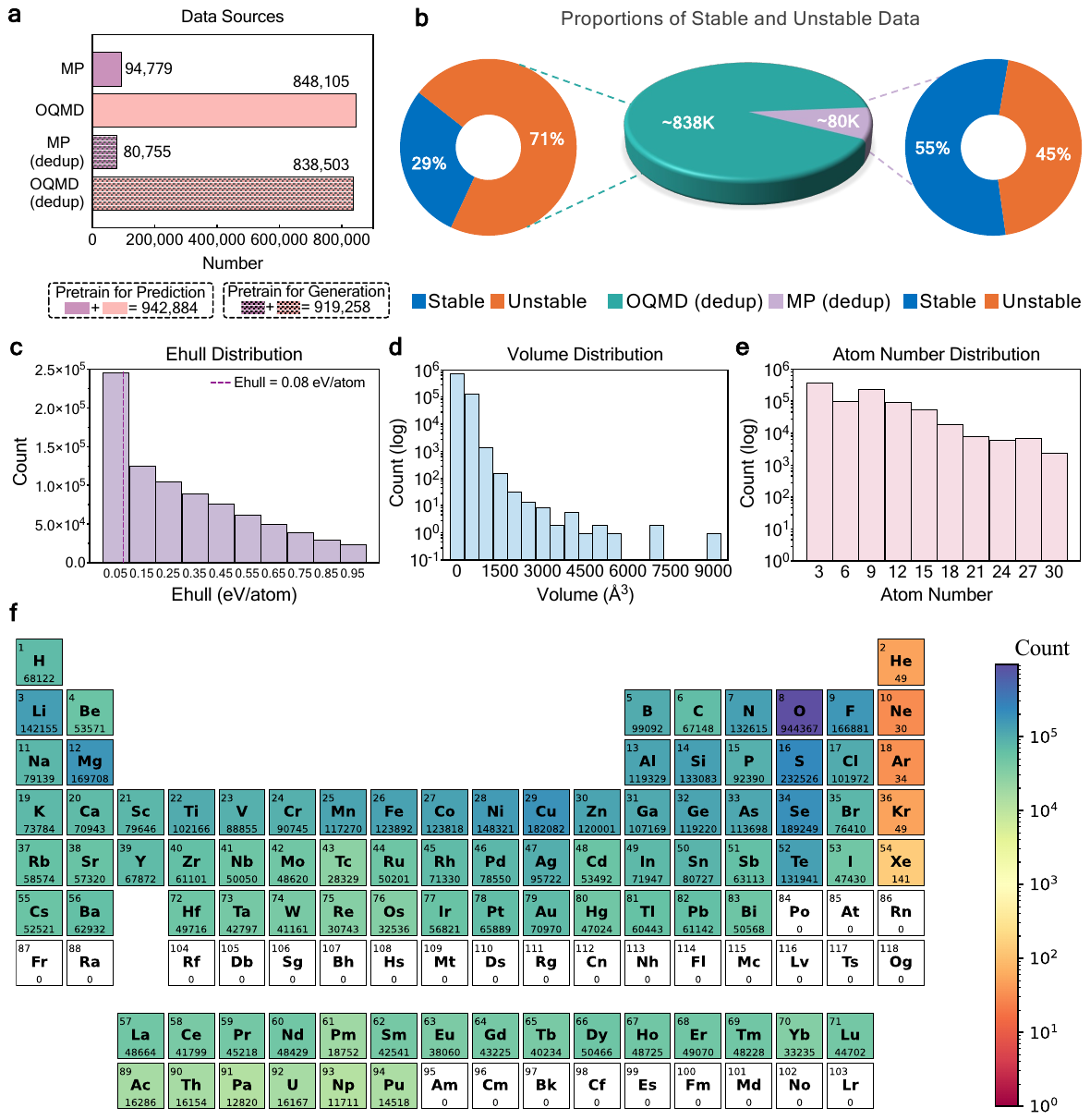}
\caption{Statistics of the pretraining dataset CrysDB. \textbf{a}, The source analyses of CrysDB and its deduplicated version from MP and OQMD. \textbf{b}, The proportions of stable and unstable structures. ``dedup'' denotes the deduplicated dataset. \textbf{c-e}, The distributions of Ehull per atom, volume and atom number per crystal, respectively. \textbf{f}, The element coverage of all crystals in CrysDB.}
\label{fig:dataset_analysis}
\end{figure}
\subsection{The Construction of CrysDB}\label{sec:pretraining_dataset}

The pretraining dataset CrysDB is sourced from two well-established open-source datasets in the field of crystal research: the Materials Project (MP)~\cite{jain2013_mp_dataset} and OQMD~\cite{kirklin2015_oqmd} datasets. Particularly, we collect crystals containing between 3 and 30 atoms, with the Ehull value below 1.0 eV/atom, yielding a total of 94,779 and 848,105 entries from MP and OQMD, respectively, as shown in~\cref{fig:dataset_analysis}a. In addition, to prevent data leakage on CSP tasks, we exclude any crystals that appear in downstream benchmarks (\emph{i.e.} MP-20 and MPTS-52), resulting in a deduplicated version of CrysDB that consists of 919,258 entries. The deduplication follows a two-step procedure. For entries with identical compositions, we use StructureMatcher from Pymatgen library~\cite{ong2013_pymatgen} to detect and remove structurally similar duplicates. Regarding entries with different compositions, structural similarity does not cause leakage because our CSP task conditions solely on composition, and no target structure information is exposed. \cref{sec:dataset_dedup} provides a comprehensive description of the dataset curation process. 
\cref{fig:dataset_analysis}b shows the proportions of stable data from OQMD (29\%) and MP (55\%). Additional statistical insights into the curated dataset are also provided. Specifically, the Ehull distribution (\cref{fig:dataset_analysis}c) reveals that the number of collected crystals 
monotonically decreases as Ehull increases. The volume distribution (\cref{fig:dataset_analysis}d) extends up to 9000\,$\text{\AA}^3$, although the majority of crystals have volumes below 1500\,$\text{\AA}^3$. The distribution of atom number per crystal (\cref{fig:dataset_analysis}e) is relatively uniform across the range of 3 to 30. Furthermore, we summarize the chemical element coverage of our dataset in~\cref{fig:dataset_analysis}f. CrysDB encompasses most groups of the periodic table, excluding heavy radioactive elements (Z=84-88 and Z=95-118), with metallic elements constituting the majority. It is important to note that these statistics presented in \cref{fig:dataset_analysis}b-f refer to the deduplicated version of CrysDB.

\subsection{Accurate Crystal Structure 
Prediction through the Finetuned DAO-G}\label{sec:exp_structure_generation}

We finetune and evaluate DAO-G on two well-recognized benchmarks for studying CSP: MP-20 (with 45,231 crystals) and MPTS-52 (with 40,476 crystals)~\cite{jain2013_mp_dataset}. Both datasets are derived from the Materials Project (MP), but they differ in complexity; MP-20 limits the number of atoms per crystal to 20, while MPTS-52 extends this to up to 52, encompassing more intricate and diverse structures.
Following DiffCSP~\cite{jiao2024diffcsp}, we utilize the same train/validation/test splits: 27136/9047/9046 for MP-20 and 27380/5000/8096 for MPTS-52, and adopt Match-Rate (MR) and Root-Mean-Square-Error (RMSE) as evaluation metrics. For each reference structure in test time, we report the best MR and RMSE of 1-shot generation. We choose the following SOTA methods for comparisons: (1) VAE-based model CDVAE~\cite{xie2022cdvae}; (2) diffusion-based models, including DiffCSP~\cite{jiao2024diffcsp}, EquiCSP~\cite{lin2024equicsp} and MatterGen~\cite{zeni2023mattergen}; (3) flow-based model, FlowMM~\cite{miller2024flowmm}.
We further conducted experiments by pretraining the baseline algorithms with the same CrysDB
data (hyperparameter settings see \cref{sec:csp_baseline_hparams}). Our implementation includes the competitively performing, open-sourced baseline models DiffCSP~\cite{jiao2024diffcsp} and FlowMM~\cite{miller2024flowmm}. These models share a nearly identical denoising architecture, though FlowMM utilizes a more advanced generative method (flow matching vs. diffusion). Our proposed DAO method adopts diffusion which is the same generative method used by DiffCSP, but DAO incorporates a more powerful Transformer-based denoising model, \emph{i.e.} Crysformer. To ensure a fair comparison, we scaled these baseline models to a parameter size comparable to Crysformer, resulting in the DiffCSP-large and FlowMM-large variants. MatterGen~\cite{zeni2023mattergen} is a pretrained model specifically designed for de novo generation (DNG) and can also be adapted for crystal structure prediction (CSP). The publicly released version was trained on a non-deduplicated version of the Materials Project dataset~\cite{jain2013_mp_dataset}. To ensure a fair comparison, we re-implement and pretrain MatterGen using the authors' official code, while scaling the model to match our own in size. Further analysis and comparison with MatterGen are presented in \cref{sec:dao_vs_mattergen}. 
To the best of our knowledge, our models are the first foundation models evaluated on these two benchmarks, leveraging the pretrain-finetune paradigm for learning from crystal structures. Notably, both DAO-G and DAO-P used in this section are pretrained on the deduplicated CrysDB.

\begin{table}[t!]
  \centering
  \setlength{\tabcolsep}{6pt} %
  \caption{CSP performance (1-shot) on the MP-20 and MPTS-52 datasets. For each metric, the best result across all methods is highlighted in bold, and the second-best result is underlined for clarity. Results for CDVAE are directly sourced from the DiffCSP paper~\cite{jiao2024diffcsp}; results for other non-pretrained baselines are taken from their original publications. All pretrained models are trained on the same pretraining dataset (CrysDB), with results averaged over three independent runs. Model sizes are also reported to facilitate fair comparison. Specifically, the ``Crysformer + DiffCSP'' configuration corresponds to our DAO-G (Stage \Rst).}
  \label{tab:structure_generation}
  \vspace{0.03in}
  \resizebox{0.98\textwidth}{!}{ 
    \begin{tabular}{ccccccc} %
    \toprule
    \multirow{2}[2]{*}{\textbf{Category}} & \multirow{2}[2]{*}{\textbf{Model}} & \multirow{2}[2]{*}{\textbf{Size}} & \multicolumn{2}{c}{\textbf{MP-20}} & \multicolumn{2}{c}{\textbf{MPTS-52}} \\
    & & & \textbf{Match Rate (\%) $\uparrow$} & \textbf{RMSE $\downarrow$} & \textbf{Match Rate (\%) $\uparrow$} & \textbf{RMSE $\downarrow$} \\
    \midrule

    \multirow{6}{*}{\makecell{\textbf{Non-} \\ \textbf{Pretrained}}} 
        & CDVAE & --   & 33.90  & 0.1045 & 5.34 & 0.2106 \\
        & DiffCSP & --    & 51.49 & 0.0631 & 12.19 & 0.1786 \\
        & EquiCSP & --  & 57.39 & 0.0510 & 14.85 & 0.1169 \\
        & FlowMM & --    & 61.39 & 0.0560 & 17.54 & 0.1726 \\
        & \makecell{Crysformer + DiffCSP} & --  & 51.55 & 0.0915 & 17.65 & 0.1428 \\ 
    \midrule
    
    \multirow{5}{*}{\textbf{Pretrained}} 
        & DiffCSP & 12.3M  & 51.23 & 0.0552 & 18.50 & 0.0825 \\
        & DiffCSP-large & 26.2M  & 64.04 & 0.0433 & 30.77 & \textbf{0.0640} \\
        & MatterGen  & 25.3M  & 67.40 & \textbf{0.0332} & 30.28 & \underline{0.0703} \\ 
        & FlowMM-large & 25.7M  & \underline{69.95} & \underline{0.0378} &  \underline{33.78} & 0.0951 \\
        & Crysformer + DiffCSP & 25.2M &65.60 & 0.0411 & 32.52 & 0.0731 \\
        & Crysformer + FlowMM  & 25.2M & \textbf{74.17} & 0.0400 & \textbf{42.01} & 0.1083 \\
        \bottomrule
    \end{tabular}%
  }
\end{table}

\begin{figure}[t!]
\centering
\includegraphics[width=0.92\textwidth]{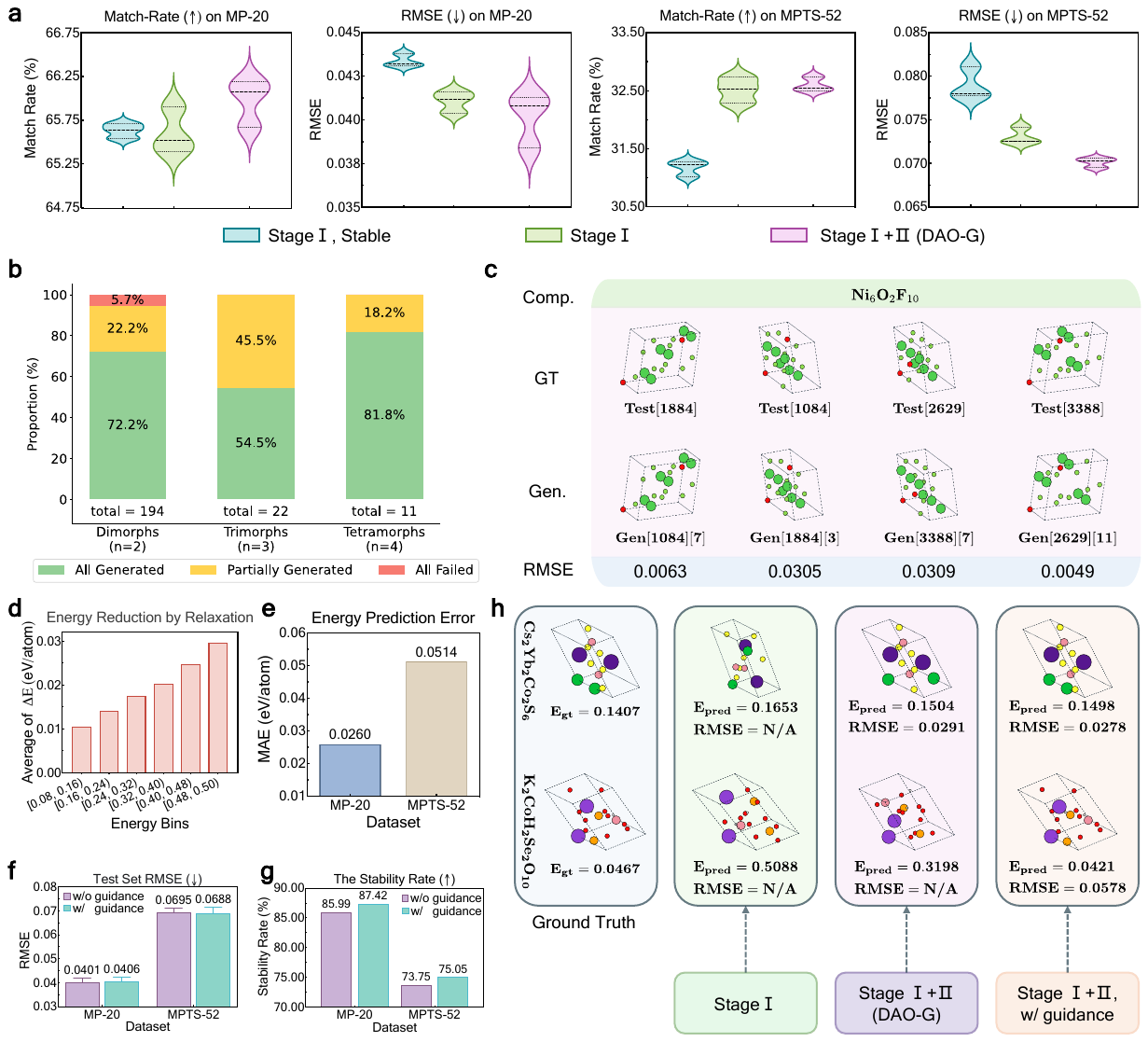}
\caption{
In-depth analyses of our models on CSP benchmarks. \textbf{a}, Performance comparison across DAO-G configurations, based on three runs, showing the median and upper and lower quartiles. ``Stage \Rst, Stable'' refers to pretraining on the stable-only subset of the deduplicated CrysDB, while ``stage \Rst'' denotes first-stage pretraining on the full deduplicated CrysDB. \textbf{b}, Proportion of structure generation outcomes across polymorphic systems. Each bar shows the distribution of three outcomes: all polymorphs successfully generated (green), partially generated (yellow), and completely failed (red). The total number of systems in each category is annotated below the bars. \textbf{c}, A 4-polymorph structure case. Abbreviations: Comp. = Composition, GT = Ground Truth, Gen. = Generation. Test[1884] denotes the 1884th test entry, and Gen[1084][7] represents the seventh of 20 generated samples based on the 5606th entry. \textbf{d}, Energy reduction after relaxation on CrysDB. $\Delta\text{E}$ is calculated using normalized energy. \textbf{e}, MAEs of energy predictions by DAO-P on the MP-20 and MPTS-52 test sets. \textbf{f} and \textbf{g}, Test RMSE and stability rate with and without energy guidance, respectively. Error bars denote standard deviation across three runs. \textbf{h}, Visualizations of two MPTS-52 examples showing the benefits of energy-guided generation. N/A denotes the failed match.}
\label{fig:generation_analysis}
\end{figure}

From the results in \cref{tab:structure_generation}, we have the following observations: 
(1) Impact of Pretraining. Firstly, large-scale pretraining leads to substantial performance gains for DAO-G, boosting the Match Rate from 51.55\% to 65.60\% on MP-20. Similarly, for DiffCSP, we observe that both pretraining and an increased model capacity result in significant improvements, further confirming the effectiveness of large-scale pretraining for crystal structure generation. With regard to FlowMM, the pretraining demonstrates a more significant performance enhancement, surpassing our original model (Crysformer + Diffusion) on the MP-20 benchmark. We attribute this to the effectiveness of its flow matching method compared to our diffusion approach. Therefore, we adopt flow matching in place of diffusion within our architecture. Interestingly, this replacement led to further gains, yielding the best Match Rates of 74.17\% on MP-20 and 42.01\% on MPTS-52. While the corresponding RMSE values remain marginally behind those of FlowMM-large, we consider Match Rate to be the primary metric, as RMSE solely reflects the error of matched structures and is  meaningful only when a high proportion of structures are successfully matched. 
These results reaffirm that our pretraining strategy and framework are generally effective and not limited to specific cases. 
(2) Efficacy of Crysformer. DAO-G can be considered a  variant of DiffCSP-large, in which the original noise prediction model is replaced with our proposed Crysformer architecture. As demonstrated empirically, DiffCSP-large is outperformed by DAO-G across nearly all evaluation metrics. A similar improvement is also evident in FlowMM (Crysformer+FlowMM vs. FlowMM-large), further validating the enhanced representational capacity of our Transformer-based denoising framework. (3) Priority on Large Systems. Comparing with MatterGen, while the retrained MatterGen performs slightly better than DAO-G on MP-20, it underperforms on the more challenging MPTS-52 dataset (i.e., MR 30.28\% vs. 32.52\%). This indicates a limitation in scaling to larger-atom systems, which is consistent with MatterGen's original training set not including structures with more than 20 atoms. More results about 20-shot sampling can be seen in \cref{sec:20_shot_results}.

\subsection{Involving Unstable Crystals is Essential in Pretraining DAO-G.}\label{sec:exp_relaxation}
In this part, we analyze the two-stage pretrained DAO-G, where DAO-P is utilized to relax the pretraining dataset from Stage \Rst \ to Stage \Rnd. Here, we conduct ablation studies to validate the benefits of this strategy. We first assess the energy prediction accuracy of DAO-P, as it is fundamental to the relaxation process. As shown in \cref{fig:generation_analysis}e, the Mean-Absolute-Errors (MAEs) of the predicted energies by DAO-P are  0.0260 eV/atom and 0.0514 eV/atom on the test sets of MP-20 and MPTS-52, respectively. Note that DAO-P is NOT even finetuned on these two benchmarks. This accuracy is typically considered acceptable for materials science, suggesting that DAO-P provides a reliable basis for structure relaxation. Besides, \cref{fig:generation_analysis}d depicts the average energy reduction across energy bins after relaxation, revealing a clear trend of energy refinement, particularly pronounced in the higher-energy bins.

We then explore the effect of involving unstable crystals for DAO-G's pretraining. For this purpose, we consider two ablations of DAO-G without the second stage: DAO-G (Stage \Rst, Stable) that use only stable crystals in deduplicated CrysDB for pretraining, and DAO-G (Stage \Rst) that is pretrained on the full deduplicated CrysDB. For clear reference, the original DAO-G is dubbed as DAO-G (Stage \Rst+\Rnd).
As visualized in \cref{fig:generation_analysis}a, after finetuning, DAO-G (Stage \Rst) outperforms DAO-G (Stage \Rst, Stable) in most cases,  suggesting that directly incorporating unstable data during pretraining enhances DAO-G’s performance. We now compare DAO-G (Stage \Rst+Stage \Rnd) with DAO-G (Stage \Rst). On MP-20, DAO-G (Stage \Rst+Stage \Rnd) achieves improved MR and lower RMSE, though with a slight increase in RMSE variance. On MPTS-52, it maintains comparable MR while significantly reducing RMSE and lowering variance for both metrics. These results highlight the effectiveness of conducting data relaxation. In \cref{sec:dao_ablation}, we conduct an ablation study on the two-stage pretraining strategy by comparing our full model against several variants, including training Stage \Rst\ with only relaxed data, removing unstable structures in the second stage to assess their impact, and employing the deduplicated-data-pretrained DAO-P model for structure generation.

\subsection{DAO-G excels in generating polymorphic structures}\label{sec:exp_poly}

In crystallography, polymorphism refers to the phenomenon in which a compound can crystallize into multiple distinct crystal structures. These polymorphs often exhibit diverse physicochemical properties, underscoring the importance of identifying the optimal structures for specific applications. Developing a generative model capable of capturing the structural diversity of polymorphs is crucial yet challenging. Such a model would need to account for the geometric variations in atomic arrangements and energetics that give rise to different polymorphic forms.
To evaluate whether DAO-G can effectively generate polymorphs, we
report some quantitative statistics, considering cases with 2-, 3-, and 4-polymorphs. For each category, we report: 1) the number of such cases in the dataset; 2) the proportion of cases where all polymorphs are successfully generated; 3) the proportion where only part of the polymorphs are recovered; 4) the proportion where none of the polymorphs are recovered. The results in \cref{fig:generation_analysis}b show that: the all-success rates are 72.2\%, 54.5\%, and 81.8\% for 2-, 3-, and 4-polymorph cases, respectively. Notably, our model successfully generates most (81.8\%) of the 4-polymorph structures, demonstrating the strong capability of DAO-G in polymorphic structure generation. The occurrence of "All Failed" cases is rare, observed only at $n$=2 with a rate of 5.7\%, and completely absent for $n$=3 and $n$=4. It is observed that the success rate does not exhibit a systematic change with an increasing number of polymorphs. We attribute this to the limited sample size available for higher-order polymorphs, which results in high variance in the estimated success rates, preventing strong conclusions about systematic behavior. Furthermore, we select a representative case from the MP-20 test set for visualization. \cref{fig:generation_analysis}c demonstrates that DAO-G successfully generates diverse conformations for the same chemical composition, achieving remarkably low RMSE values in each case. 
For the case of $\text{Ni}_6\text{O}_2\text{F}_{10}$, which has 4 distinct conformations in the test set, we generate 20 samples for each conformation, resulting in a total of 80 ($=20\times 4$) generated structures. We can observe that all the 4 different conformations are hit successfully by the generated structures, with correspondingly low RMSE values of 0.0063, 0.0305, 0.0309, and 0.0049. This indicates the model's ability to accurately capture and replicate the structural diversity of polymorphic systems. More visualizations see \cref{sec:Polymorph_more_vis}.

\subsection{Energy-Guided Sampling by DAO-P Contributes to Higher-Stability Generation}\label{sec:exp_guidance}

As outlined in \cref{sec:pretrain_finetune_framework}, DAO-P is pretrained to predict the intermediate energy of noise-added crystals along the diffusion path. To harness this capability, we apply DAO-P as the energy guider and implement energy-guided sampling for the generation process of DAO-G, following the methodology introduced in~\cite{jiao2024diffcsp}. Further details are provided in~\cref{sec:energy_guidance}. Here, we conduct quantitative analyses from three perspectives. First, we report the test RMSE on MP-20 and MPTS-52 in \cref{fig:generation_analysis}f. It is observed that incorporating energy guidance yields comparable RMSE on MP-20, but a lower RMSE ($0.0695 \to 0.0688$) on the more difficult dataset MPTS-52, suggesting that energy guidance is particularly beneficial for generating complex crystal structures. Second, we examine the stability rate of the generated structures in \cref{fig:generation_analysis}g. Clearly, DAO-G (w/ guidance) exhibits higher stability rates compared to the counterpart without guidance on both MP-20  and MPTS-52, namely, 87.42\% vs. 85.99\% and 75.05\% vs. 73.75\%, respectively. It confirms our hypothesis that energy guidance does promote structural stability. Finally, we visualize two specific examples in \cref{fig:generation_analysis}h to provide further insight into the impact of energy guidance. For the first example, $\text{Cs}_2\text{Yb}_2\text{Co}_{2}\text{S}_6$, while DAO-G achieves relatively low RMSE and energy values compared to the one-stage model, the incorporation of energy guidance further enhances performance. Specifically, the RMSE decreases from 0.0291 to 0.0278, and the predicted energy is reduced from 0.1504 eV/atom to 0.1498 eV/atom.
For the second example, we choose a considerably  more complex crystal, $\text{K}_2\text{Co}_2\text{H}_4\text{Se}_2\text{O}_{10}$. In this case, only the energy-guided DAO-G successfully matches the ground truth, as demonstrated by its low RMSE of 0.0578 and its precise energy prediction relative to the ground truth (0.0421 vs. 0.0467). With the inclusion of energy guidance, DAO-G achieves a significant energy reduction of 86.8\%, decreasing from 0.3198 to 0.0421. The results above support the critical role of energy guidance in handling complex crystal structures and achieving both structural and energetic accuracy.

\begin{figure}[t!]
\centering
\includegraphics[width=0.88\textwidth]{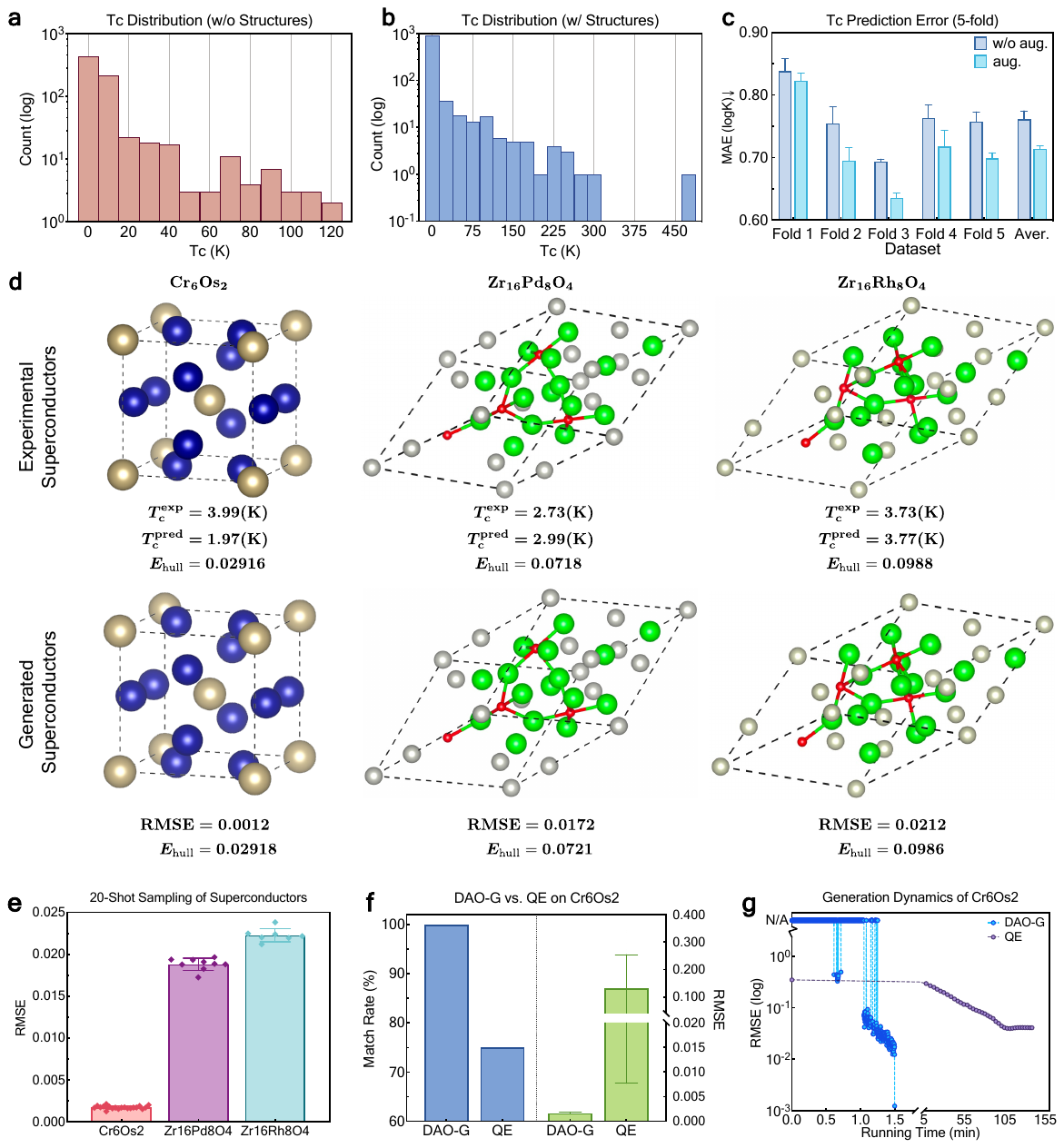}
\caption{Experiments on superconductors. \textbf{a-b}, Critical temperature distributions for crystals with and without structures in the SuperCon3D~\cite{chen2025supercon3d}. \textbf{c}, $T_c$ prediction error comparison between DAO-P trained on original and augmented structures (aug.) using 5-fold cross-validation. Error bars represent the standard deviation from three independent runs. \textbf{d}, Results on three real-world superconductors. The top row shows experimental structures, critical temperatures and Ehull values (DFT), while the bottom row displays DAO-G generated structures, corresponding RMSEs, and Ehull values. \textbf{e}, RMSEs of 20-shot DAO-G sampling for the three superconductors. Error bars indicate RMSE standard deviation of successfully matched samples (20, 9, and 7 for each superconductor). \textbf{f}, Comparison of DAO-G and QE optimizer in terms of MR and RMSE for $\text{Cr}_6\text{Os}_2$. Error bars denote the same as in panel \textbf{e}, with 20 and 5 samples. \textbf{g}, Sampling dynamics of DAO-G and optimization dynamics of QE optimizer for $\text{Cr}_6\text{Os}_2$. N/A denotes failed matches.}
\label{fig:supercon}
\end{figure}

\subsection{DAO-G and DAO-P Demonstrate Significant Potential for Superconductor Analysis}\label{sec:exp_supercon}

In this section, we evaluate the capabilities of DAO-G and DAO-P in analyzing superconductors. Specifically, we focus on two key tasks: the prediction of 3D crystal structures and the estimation of the critical temperature $T_c$. It is noteworthy that predicting the structure of superconductors is more complex than that of ordinary crystals due to their intricate structural features and the sensitivity of superconductivity to structural variations. Determining the critical temperature presents an even greater challenge. For conventional superconductors, this requires the application of Bardeen-Cooper-Schrieffer (BCS) theory~\cite{bardeen1957bcs} to model electron-phonon interactions. For unconventional superconductors, other theoretical models are necessary. These complexities highlight the need for DAO-G and DAO-P, which can address these challenges through their effectiveness in structure generation and property prediction.

We finetune our pretrained models using the publicly available SuperCon dataset~\cite{supercon,chen2025supercon3d}, which includes 1017 ordered superconductors with known structures and 748 ordered superconductors without structural information. \cref{fig:supercon}a,b provide a visualization of $T_c$ distribution for both categories, exhibiting a broad range of values. We first finetune DAO-G on the subset of entries with structures and then use it to generate structures (1-shot sampling) for the remaining 748 compositions to augment the original dataset. Finally, we finetune DAO-P to predict $T_c$ using the original dataset and its augmented version, respectively, under the 5-fold cross-validation setting (see \cref{sec:supercon3d_aug}). \cref{fig:supercon}c shows that augmenting DAO-P with generated structures consistently improves performance across all 5 folds, reducing the average MAE (logK) from 0.761 to 0.714. These results confirm the strong potential of DAO-G in practical scenarios where structural information is unavailable and demonstrate that incorporating the structures generated by DAO-G can enhance the performance of DAO-P.

To validate the practicability of our models on real-world systems, we further conduct evaluations on three real-world superconductors: $\text{Cr}_6\text{Os}_2$~\cite{flukiger1974electronically}, $\text{Zr}_{16}\text{Pd}_8\text{O}_4$~\cite{watanabe2023_sc2} and $ \text{Zr}_{16}\text{Rh}_8\text{O}_4$~\cite{watanabe2023_sc2}.
These three superconductors are unseen during both the pretraining and finetuning processes, thereby avoiding the risk of data leakage. The ensemble of the five DAO-P models, finetuned under the aforementioned 5-fold cross-validation setting, is leveraged to predict the $T_c$ values for the three structures. Remarkably, the absolute errors remain within a modest range, being 2.02 (K), 0.26 (K) and 0.04 (K), as shown in \cref{fig:supercon}d.

In addition, we employ the finetuned DAO-G to generate crystal structures for each superconductor and present the most accurate ones of 20-shot sampling in \cref{fig:supercon}d. To further validate the stability of the structures generated by DAO-G, we perform DFT calculations to evaluate the energy above hull (Ehull) for each generated material. Since the Phase Diagram data provided by MP~\cite{jain2013_mp_dataset} is based on VASP~\cite{hafner2008ab} with specific computational settings, we used the same tool and parameters to ensure strict consistency when calculating the total energies. Determining the structures of these three superconductors is known to be challenging. For instance, $\text{Cr}_6\text{Os}_2$ crystallizes in the A15 structure, where Cr atoms form a three-dimensional network of linear chains and Os atoms occupy the body-centered cubic sites. DAO-G achieves a 100\% MR and impressively low RMSE of 0.0012. Notably, both MP and OQMD~\cite{jain2013_mp_dataset,kirklin2015_oqmd} (the pretraining datasets) contain multiple entries with the same chemical formula but distinct crystal structures. Among them, some structures are stable (Ehull $\approx 0.026$) and others are unstable (Ehull $> 0.08$). Importantly, the stable entries were explicitly removed from our pretraining data during the deduplication process against the MP-20 and MPTS-52 benchmarks. Interestingly, although the model encountered unstable $\text{Cr}_6\text{Os}_2$ structures during pretraining, it does not show a bias toward reproducing them. Instead, DAO-G generates the stable superconducting structure, with a DFT-calculated Ehull of 0.02918, closely matching the experimental value (0.02916). This indicates that DAO-G is not merely memorizing training examples but has learned a meaningful distribution over stable crystal structures conditioned on composition, even when unstable polymorphs exist in the training data. For the second superconductor $\text{Zr}_{16}\text{Pd}_8\text{O}_4$, it exhibits $\eta$-carbide structure (space group $Fd\bar{3}m$) with intricate atomic arrangements, such as rigid Wyckoff site occupancy and geometrically frustrated stella quadrangula lattices. DAO-G successfully generates the structure with an RMSE of 0.0172. In contrast,  $\text{Zr}_{16}\text{Rh}_8\text{O}_4$ exhibits a minor change in lattice constant ($\sim 0.5\%$) when substituting Rh for Pd, which, yet, significantly affects its superconducting properties, such as increasing $T_c$ from 2.73 K to 3.73 K. Remarkably, DAO-G is still capable of resolving this subtle structural change, achieving a low RMSE of 0.0212. 
For the three superconductors, the Ehull values of the generated structures are very close to those of the experimental ones, with differences of only 0.00002, 0.0003, and 0.0002 eV/atom, respectively. This demonstrates that the accuracy of DAO-G is not only reflected in the RMSE of structural reconstruction but is also robustly confirmed in terms of thermodynamic stability. \cref{fig:supercon}e records RMSEs of 20-shot sampling for all the three superconductors, where the variances are observed to be minor in all cases, demonstrating the strong robustness of our model.  

We also try to derive the structures of the three superconductors using the DFT-based structure predictor based on the Quantum Espresso (QE) optimizer~\cite{giannozzi2017qe}. We initialize the structures by adding random noise to the experimental structures. QE optimizer only yields reasonable structures for 
$\text{Cr}_6\text{Os}_2$, failing to do so for the other two superconductors. In~\cref{fig:supercon}f, across 20 independent runs, QE optimizer achieves a MR of 75\% and an average RMSE of 0.1310, with respect to the experimental structures. Importantly, in our setting, unreasonable structures are defined as those that fail to match ground-truth crystals according to StructureMatcher~\cite{ong2013_pymatgen}. The gap between the DFT-relaxed structure and the ground truth, despite a relatively small RMSE, can be explained by two factors: First, we introduced slight perturbations to the experimental structures before relaxation to demonstrate the robustness and practical advantages of DAO-G under less informative or more challenging initial conditions. Such perturbations could lead DFT to converge to a nearby local minimum. Second, DFT relaxations represent 0 K equilibrium structures, whereas experimental data are obtained at finite temperatures (e.g., 3.99 K for $\text{Cr}_6\text{Os}_2$~\cite{flukiger1974electronically}), inevitably introducing discrepancies even for highly accurate calculations. Although global optimization frameworks like USPEX~\cite{glass2006uspex} and CALYPSO~\cite{wang2012calypso} combine DFT with evolutionary search to reduce local-minimum trapping, they require substantially higher computational cost. Even when accelerated by fast interatomic potentials (e.g., FUSE~\cite{collins2025fuse} and AIRSS~\cite{pickard2025airss}), their performance strongly depends on the accuracy of the predicted potential; errors that accumulate during relaxation can lead to significant deviations in the final stable structure. This approach, similar to but distinct from methods employing foundation potentials, directly learns the distribution of stable crystal structures, thereby achieving a favorable balance between competitive efficiency and high accuracy.
Furthermore, \cref{fig:supercon}g visualizes the sampling dynamics of DAO-G and the optimization dynamics of QE optimizer. Notably, even under more advantageous initialization, where the initial RMSE to the ground truth is finite and nontrivial, DFT remains significantly slower, where QE takes about 138 minutes over 38 iterations, whereas our DAO-G completes 1000 sampling iterations in just 1.5 minutes. In other words, the computation speed per iteration of QE optimizer
is more than 2000 times slower than DAO-G. Comprehensive details are provided in \cref{sec:supercon_app}.

\section{Discussion}\label{sec:discussion}
This paper presents Siamese foundation models specifically designed to address CSP challenges. Our pretrain-finetune framework, named DAO, comprises two complementary foundation models: DAO-G for structure generation and DAO-P for energy prediction, both of which are built upon a proposed expressive graph Transformer---Crysformer, and are pretrained using meticulously crafted strategies. For the pretraining, we construct a crystal pretraining dataset (i.e., CrysDB) of about 940K entries with structures and the corresponding energy labels. A key feature of our approach is the cooperative interaction between DAO-G and DAO-P: DAO-P enhances DAO-G's performance by utilizing unstable data through a dedicated relaxation process and energy-guided sampling, while DAO-G generates augmented structures for DAO-P when structural information is unavailable in practice. Notably, the majority of existing CSP models are trained on relatively small datasets such as MP-20 and MPTS-52, which limits their ability to learn diverse structural patterns and generalize to more complex systems. By contrast, the large-scale, high-quality CrysDB dataset assembled in this work enables effective pretraining, providing the model with broader structural knowledge and significantly improving its generative performance and robustness.

We rigorously evaluate the performance of our method across a variety of downstream tasks. In CSP evaluation, DAO-G achieves SOTA results on both MP-20 and MPTS-52 datasets, highlighting its ability to generate stable crystal structures accurately. Ablation studies validate the benefits of our proposed dataset relaxation and energy guidance for improving the generative performance. Meanwhile, DAO-G is proven effective at generating polymorphic structures. Moreover, DAO-P demonstrates exceptional transferability in property prediction tasks, outperforming existing methods on four out of eight datasets. It is also worth mentioning that experiments on superconductors provide further evidence of the synergistic interaction and mutual performance enhancement between DAO-G and DAO-P, demonstrating the practicability of our models on real-world systems.

Although our approach has demonstrated the significant potential of foundation models for advancing materials science research and development, several challenges remain to be addressed to further enhance practicality. First, the performance on the MPTS-52 benchmark, particularly regarding the match rate (reaching only 46.78\% in the 20-shot setting), remains unsatisfactory. MPTS-52 includes crystals with up to 52 atoms, whereas our pretraining dataset is limited to crystals with atom numbers ranging from 3 to 30. We hypothesize that expanding the pretraining dataset to include more large-sized crystals could significantly improve the model's performance. Second, while we employ a diffusion model to define the CSP pretraining objective, other generative models, such as flow matching, have shown promise in CSP tasks. Exploring the integration of more advanced generative models into the pretraining process represents a promising direction for future work. Finally, in our experiments on superconductor analysis, we focus on structure generation and critical temperature prediction, but a more meaningful task would be the design of novel high-temperature superconductors, which remains a challenging and underexplored area. Moving forward, we plan to leverage our models to screen existing superconductors and enable property-guided design of novel high-temperature superconductors.

\section{Methods}\label{sec:methods}

\subsection{Crystal Formulation}
A crystal possesses a repeating structural unit that extends infinitely throughout 3D space. The unit comprises a lattice and atoms situated within it. The periodicity of a crystal lattice is embodied in three vectors, $\vec{\rmL}=[\vec{\rvl}_1, \vec{\rvl}_2, \vec{\rvl}_3]\in\sR^{3\times 3}$, each defining a direction along which the structure repeats. Within the lattice, the chemical composition is  characterized as atom types $\rmA=[\rva_1, \rva_2, \dots, \rva_N]\in\sR^{1\times N}$, and the atom Cartesian coordinates are denoted as $\vec{\rmX}=[\vec{\rvx}_1, \vec{\rvx}_2, \dots, \vec{\rvx}_N]\in\sR^{3\times N}$, where $N$ is the number of atoms. Fractional coordinates, $\rmF=\vec{\rmL}^{-1}\vec{\rmX}=[\rvf_1, \rvf_2, \dots, \rvf_N]\in [0, 1)^{3\times N}$, offer an alternative format to represent atom locations. This representation inherently places atoms within the lattice, which is particularly useful for crystal structure generation. We employ fractional coordinates in this paper, as opposed to Cartesian coordinates. As a result, a crystal can be shown as $\gM=(\rmA, \vec{\rmL}, \rmF)$.

\subsection{Deduplication of the Pretraining Dataset}
\label{sec:dataset_dedup}
When pretraining DAO-G, we curated the pretraining dataset by removing any samples included in MP-20 and MPTS-52 test sets, thus mitigating the risk of data leakage and ensuring a robust evaluation. The procedure is introduced in \cref{alg:dedup}. Notably, more stringent thresholds of \emph{StructureMatcher}~\cite{ong2013_pymatgen} are employed here than in the generation performance evaluation.

For crystals with similar structures, two scenarios are possible: (1) Same chemical composition. Our data deduplication employs StructureMatcher to compute structural similarity. When a match is found between entries of the same composition, the duplicate is removed, ensuring all such similar structures are eliminated from the pretraining data. (2) Different chemical compositions. This scenario does not constitute data leakage. Our task is CSP, where the model learns to predict a structure from a composition. Although the pretraining dataset may contain similar structures, the model receives different compositions as input, so no information about the target structure is leaked. Therefore, based on our deduplication procedure, data leakage does not occur in our experiments.

\begin{algorithm}
\caption{Deduplication Process of Pretraining Dataset CrysDB}
\label{alg:dedup}
\begin{algorithmic}[1]
\State \textbf{Input:} Pretraining Dataset $\sD$, Test Set $\sT$
\State \textbf{Output:} Deduplicated Pretraining Dataset $\sD_{\text{dedup}}$
\State \textbf{Define:} SM = Structure-Matcher(stol=0.1, angle\_tol=10, ltol=0.3)  \Comment{Smaller stol}
\State $\sD_{\text{dedup}} \gets \emptyset$
\For{each sample $\gM_d$ in $\sD$}
    \If{$\gM_d.\text{id} \in \sT.\text{ids}$}
        \State \textbf{continue} \Comment{Skip samples in the test set}
    \EndIf
    \State $\sS \gets \{ \gM \in \sT \,|\, \gM.\text{formula} = \gM_d.\text{formula} \}$ \Comment{Find samples with the same formula}
    \State overlap-flag $\gets$ False
    \For{each sample $\gM$ in $\sS$}
        \If{$\text{SM}(\gM_d.\text{structure}, \gM.\text{structure})$} \Comment{Same formula and similar structures}
            \State $\text{overlap-flag} \gets \text{True}$ \Comment{Mark as overlap}
            \State \textbf{break} \Comment{No need to check further}
        \EndIf
    \EndFor
    \If{\textbf{not} overlap-flag}
        \State $\sD_{\text{dedup}} = \sD_{\text{dedup}} \cup \{\gM_d\}$ \Comment{Add it into the output}
    \EndIf
\EndFor
\State \textbf{Return} $\sD_{\text{dedup}}$ \Comment{Return the deduplicated dataset}
\end{algorithmic}
\end{algorithm}

\subsection{Equivariant Diffusion Models for Crystal Structure Generation}\label{sec:diffusion_training}
Inspired by DiffCSP~\cite{jiao2024diffcsp}, our diffusion-based pretraining can be formulated as $p(\vec{\rmL}, \rmF|\rmA)$, that is, generating lattice and fractional coordinates given the chemical composition. This distribution needs to adhere to $\mathrm{O}(3)$ invariance and periodical translation invariance, which for simplicity we collectively refer to as OP-invariance. It suggests that lattice rotation and coordinate translation do not affect the generation process. The notation of $\gM_t$ denotes a crystal at timestep $t$ in the diffusion process. A typical diffusion model consists of a forward process and a backward process.

In the forward process, different noise schedules are adopted for lattice and fractional coordinates. Particularly, we employ a noise addition strategy used in DDPM~\cite{ho2020ddpm} for the lattice, which is modeled as: 
\begin{equation}
q(\vec{\rmL}_t|\vec{\rmL}_0)=\gN\left(\vec{\rmL}_t|\sqrt{\bar{\alpha}_t}\vec{\rmL}_0, (1-\bar{\alpha}_t)\rmI\right), \ \ \bar{\alpha}_t=\prod\limits_{s=1}^t \alpha_s, 
\end{equation}
where $\alpha_s$ is scale of the noise added at timestep $s$. 

In contrast, for fractional coordinates, we employ a Score-Matching method~\cite{song2021maximum} and instantiate the forward process using the Wrapped Normal (WN) Distribution~\cite{de2022riemannian}:
\begin{equation}
q(\rmF_t|\rmF_0)\propto\sum_{\rmZ\in\mathbb{Z}^{3\times N}}\exp\left(\frac{-{\left\|\rmF_t-\rmF_0+\rmZ\right\|_{\text{F}}^2}}{{2 \gamma_t^2}}\right),
\end{equation}
where, $\gamma_t$ is sampled from an exponential scheduler. The use of WN distribution provides a significant advantage in reflecting the periodic nature of crystal structures~\cite{jiao2024diffcsp}. 

During the backward process, we seek to reconstruct the added noise $\bm\epsilon^L_t$ for lattice, and reconstruct the score for fractional coordinates. The corresponding training objectives are implemented as:
\begin{align}
\gL_L&=\E_{\bm\epsilon_t^L\sim\gN(0,\rmI), t\sim \gU(1,T)}\left[\left\|\hat{\bm{\epsilon}}_L(\gM_t, t)-\bm\epsilon^L_t\right\|_2^2\right],  \label{equ:loss_L}\\
\gL_F&=\E_{\rmF_t\sim q(\rmF_t|\rmF_0), t\sim \gU(1,T)}\left[\lambda_t\left\|\hat{\bm{\epsilon}}_F(\gM_t, t)-\nabla_{\rmF_t}\log q(\rmF_t|\rmF_0)\right\|_2^2\right],  \label{equ:loss_F}
\end{align}
where, $\hat{\bm{\epsilon}}_L(\gM_t, t)$ is the predicted noise for lattice,  $\hat{\bm{\epsilon}}_F(\gM_t, t)$ is the predicted score for coordinates and $\lambda_t$ is estimated by Monte-Carlo sampling. For the pretraining of DAO-G, the training objective is the combination of $\gL_L$ and $\gL_F$.

Importantly, to achieve the distribution OP-invariant, $\hat{\bm\epsilon}_L$ is supposed to be $\mathrm{O}(3)$ equivariant and $\hat{\bm\epsilon}_F$ needs to satisfy periodical translation invariance. In the following section (\cref{sec:crysformer}), we introduce a powerful graph Transformer to output the noises $\hat{\bm\epsilon}_L$ and $\hat{\bm\epsilon}_F$, which retains these symmetries, capable of effectively capturing the intricacies of crystal structures and accomplishing the denoising process.

\subsection{Equivariant Graph Transformer: Crysformer}\label{sec:crysformer}

\begin{figure}[t!]
\centering
\includegraphics[scale=0.85]{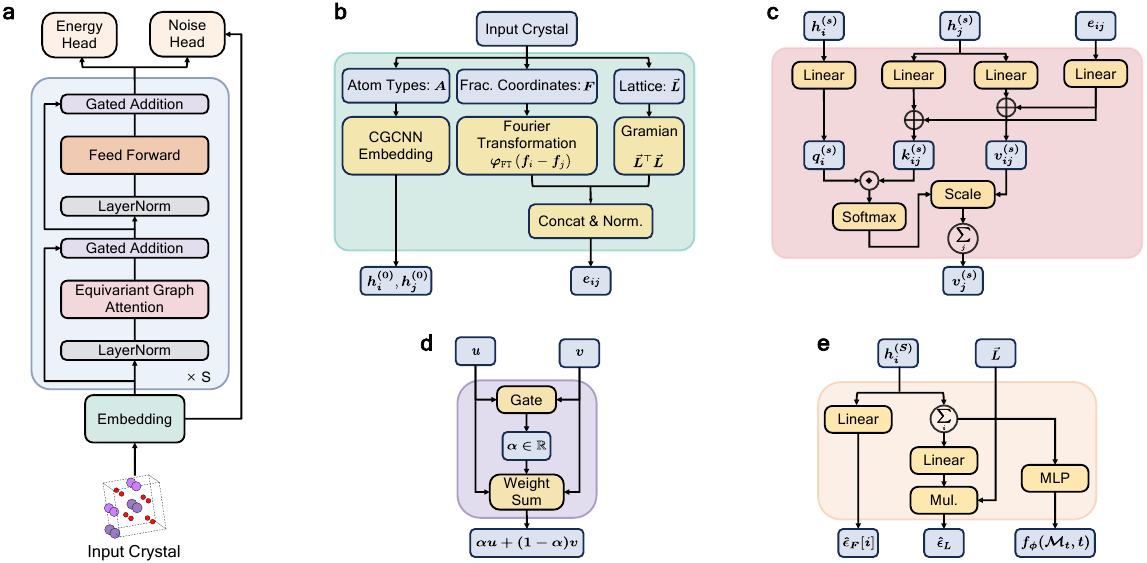}
\caption{\textbf{a}, Schematic of the Crysformer architecture. \textbf{b-e}, Detailed view of key modules within Crysformer: \textbf{b}, The embedding module generates initial node and edge representations. ``Frac.'' denotes Fractional; \textbf{c}, The invariant attention module captures interactions between nodes in an $\mathrm{O}(3)$ and periodically invariant manner; \textbf{d}, The gated addition module facilitates residual connections to improve training stability; \textbf{e}, Noise and energy prediction heads are used for denoising and energy prediction, respectively. ``Mul.'' denotes Multiplication. Note that DAO-G utilizes only the noise head, while DAO-P employs both.}
\label{fig:crysformer}
\end{figure}

Recognizing the proven power of the Transformer architecture in pretraining~\cite{vaswani2017_Transformer,luo2022_transformer-m}, we adopt graph Transformer (\cref{fig:crysformer}) as our foundational building block to replace the GNN-based architecture used in DiffCSP~\cite{jiao2024diffcsp}. Our Crysformer mainly consists of four modules: an embedding module for initializing node and edge features; an invariant graph attention module for capturing inter-node interactions; a gated addition module for residual connections; and the noise and energy prediction heads for outputting predictions. Subsequently, we present a comprehensive description of each module.

As shown in \cref{fig:crysformer}b, we firstly initialize the node features $\rmH^{(0)}=[\rvh_1^{(0)}, \rvh_2^{(0)}, \dots, \rvh_N^{(0)}]\in\sR^{d_h \times N}$ with CGCNN embeddings~\cite{xie2018cgcnn}, where $d_h$ denotes the node feature dimension. Prior to feeding the crystal into the model, we precompute edge features, represented as $\rmE=\{\rve_{ij}\} \in \sR^{d_e \times N_e}$. $d_e$ and $N_e$ are the dimension of edge features and the number of edges, respectively. Specifically, the edge feature between atom $i$ and $j$ is calculated by:
\begin{equation}
    \rve_{ij}=\texttt{Normalization}\left(\vec{\rmL}^\top \vec{\rmL} \ \| \ \bm\varphi_{\texttt{FT}}\left(\rvf_{i} - \rvf_{j}\right)\right),
\end{equation}
where $\|$ denotes concatenation and $\bm\varphi_{\texttt{FT}}$ is Fourier Transform (FT) to ensure the edge feature satisfy translation invariant following DiffCSP~\cite{jiao2024diffcsp}.

The way we use edge features in our Crysformer differs from DiffCSP. While DiffCSP employs a GNN-based backbone with an EGNN-like design to ensure symmetry, we instead implement an Equivariant Graph Transformer, which provides stronger generalization and expressive power. In DiffCSP, the results of the Fourier Transformations are concatenated with the node features and then passed through the same neural network to compute the messages between nodes. In contrast, our method applies separate neural networks to node and edge features: the networks $\bm\varphi_k,\bm\varphi_v$ are used for the keys and values, while $\bm\varphi_e$ is dedicated to the edge features. The feature maps produced by $\bm\varphi_e$ are subsequently added to the outputs of $\bm\varphi_k$ and $\bm\varphi_v$, respectively, in a corresponding element-wise fashion. We design the $s$-th attention layer (\cref{fig:crysformer}c) as $\rmZ^{(s)}=\texttt{ATT}(\rmH^{(s-1)}, \rmE) = [\rvz_{0}^{(s)}, \rvz_{1}^{(s)},\dots,\rvz_{N}^{(s)}]$, with four parametric MLPs: $\bm\varphi_q, \bm\varphi_k, \bm\varphi_v$ and $ \bm\varphi_e$. The details are as follows:
\begin{equation}
    \begin{rcases}
        \rvq^{(s)}_{i} &=\bm\varphi_q(\rvh_{i}^{(s-1)})\\
        \rvk^{(s)}_{ij} &= \bm\varphi_k(\rvh_{j}^{(s-1)}) + \bm\varphi_e(\rve_{ij})\\
        \rvv^{(s)}_{ij} &= \bm\varphi_v(\rvh_{j}^{(s-1)}) + \bm\varphi_e(\rve_{ij})
    \end{rcases}\implies
    \rvz_{i}^{(s)} =\sum_{j=1}^N \softmax_j\left(\frac{\langle\rvq^{(s)}_i, \rvk^{(s)}_{ij}\rangle}{\sqrt{d_h}}\right) \rvv^{(s)}_{ij}.
\end{equation}
In implementation, we use multi-head attention in our model, in line with the standard Transformer~\cite{vaswani2017_Transformer}. 

We also utilize residual connection in the model design. A primary difference arises from the using of Gated Residual (GR) connection (\cref{fig:crysformer}d), written as:
\begin{equation}
\texttt{GR}(\rvx, \bm\psi(\rvx))=\alpha \rvx + (1-\alpha) \bm\psi(\rvx), \ \ \ \alpha=\bm\varphi_g(\rvx, \bm\psi(\rvx), \rvx-\bm\psi(\rvx))\in\sR,
\end{equation} 
where $\bm\psi(\rvx)$ is a function with respect to $\rvx$ and $\bm\varphi_g$ is a MLP used to learn the gating coefficient $\alpha$. Compared to the vanilla residue connection~\cite{he2016resnet}, gated addition provides a gate mechanism to control the strengths of the input and layer output, which is more flexible and robust. 

Moreover, we adopt PreNorm (PRN) layer normalization in our architecture~\cite{xiong2020layer_norm}, as opposed to the PostNorm employed in the vanilla Transformer design. With the aforementioned several techniques, we can build a Crysformer block as:
\begin{align}
    \rmZ^{(s)}&=\texttt{GR}\left(\rmH^{(s-1)}, \texttt{ATT}\left(\texttt{PRN}\left(\rmH^{(s-1)}\right), \rmE\right)\right), \\
    \rmH^{(s)}&=\texttt{GR}\left(\rmZ^{(s)}, \texttt{FFN}\left(\texttt{PRN}\left(\rmZ^{(s-1)}\right)\right)\right).
\end{align}
Here, $\texttt{FFN}$ is the feedforward layer. Our Crysformer is a stack of $S$ blocks, where $\rmH^{(S)}$ contains node features extracted from the final layer. We conduct average pooling of the node features to obtain the crystal feature $\bar{\rvh}=\frac{1}{N}\sum_{i=1}^N \rvh_i^{(S)}$. 
The notations $\bar{\rvh}_t$ and $\rmH^{(S)}_t$ denote the crystal feature and the set of node features at timestep $t$, with $t=0$ by default if not explicitly specified. The node feature $\rvh_i^{(S)}$ is designed to be $\mathrm{O}(3)$-invariant, and the proof is provided in \cref{sec:model_invariance_proof}. 

Then, the learned node and crystal features can be used to predict both the lattice noise $\hat{\bm{\epsilon}}_L(\gM_t, t)$ and the fractional coordinates score $\hat{\bm{\epsilon}}_F(\gM_t, t)$. Concretely, as illustrated in Extended Data \cref{fig:crysformer}e, we feed the node features into a neural network $\bm\varphi_F$ to predict the fractional coordinates score, through $\hat{\bm{\epsilon}}_F(\gM_t, t)[i,:]=\bm\varphi_F(\rmH^{(S)}_t[i, :])$, where $[i,:]$ selects the $i$-th atom. Since the lattice belongs to the entire crystal, we use the crystal feature as input to another neural network $\bm\varphi_L$. We further multiply the output of $\bm\varphi_L$ by the lattice to make sure that the predicted noise meets $\mathrm{O(3)}$-equivariance, yielding $\hat{\bm{\epsilon}}_L(\gM_t, t)=\vec{\rmL}\bm\varphi_L(\bar{\rvh}_t)$. Meanwhile, for the energy prediction, we implement the energy head $\bm\varphi_E$ with a two-layer MLP and output the energy using $\bar{\rvh}_t$ as the input, that is, $\bm\varphi_E(\bar{\rvh}_t)$.  More details are shown in \cref{sec:noise_equivariance_proof}.

\subsection{Two-Stage Pretraining for DAO-G}\label{sec:two_stage_pretraining}
We now provide a detailed explanation of each stage involved in DAO-G pretraining and the corresponding dataset relaxation procedure.

The first stage of our pretraining strategy involves training DAO-G on the full CrysDB (dedup), which encompasses both stable and a considerable proportion of unstable crystals. This broad training dataset, combined with the diffusion process proposed by DiffCSP~\cite{jiao2024diffcsp}, allows DAO-G to learn a more generalized representation of crystal structures, capturing the variability present in both stable and unstable crystals. Specifically, different noise schedules are adopted (\cref{sec:diffusion_training}), where we choose the standard DDPM~\cite{ho2020ddpm} for lattice generation, while for fractional coordinates we employ a Score-Matching method~\cite{song2021maximum} and instantiate the forward process using the Wrapped Normal (WN) Distribution~\cite{de2022riemannian}. During training, DAO-G learns to predict the lattice noise $\epsilon_L(\gM_t, t)$ and fractional coordinate score $\epsilon_F(\gM_t, t)$, guided by the loss functions defined in~\cref{equ:loss_F,equ:loss_L}. By pretraining DAO-G on both stable and unstable data in the first stage, we aim to capture a wider distribution of crystal structures.

Despite the benefits of incorporating unstable data, employing them as input can bias the generation towards energetically unfavorable regions of the energy landscape. A straightforward idea is to relax the unstable data for quality improvement using DFT~\cite{kohn1965dft}, which is a widely used and reliable method for relaxing biochemical structures, but its high computational cost presents a significant challenge. As a more efficient alternative, we propose a machine learning-based approach for the relaxation of unstable crystals. Specifically, using the pretrained DAO-P, we can predict the energy $f_\phi(\gM, 0)$ for a given crystal $\gM$ and compute the corresponding structural energy gradient (i.e., $\nabla_{\vec{\rmL}} f_\phi(\gM, 0)$ and $\nabla_{\rmF} f_\phi(\gM, 0)$), which are subsequently used as input for L-BFGS~\cite{liu1989lbfgs} to guide the optimization process towards a minimal energy state. Actually, the L-BFGS optimization is performed using the PyTorch library~\cite{paszke2019pytorch} with the following hyperparameter configurations: max\_iter = 5, lr = 1.0. Notably, we only relax data whose Ehull ranges from 0.08 eV/atom to 0.5 eV/atom. Here, the choice of relaxation algorithm is flexible, and other similar gradient-based optimization methods are also suitable.

After relaxation, we combine the remaining original data ($\text{Ehull}\in[0,0.08)\cup (0.5,1.0]$ eV/atom) with the relaxed data to form a new pretraining dataset, whose size is same to the original one. In the second stage, we refine DAO-G by continuing training on the relaxed dataset. Using this dataset and a reduced learning rate, we resume training the model initialized with the parameters from the first stage and employ the same training losses. This allows for a refinement of the denoising process based on the improved data quality.

The hyperparameters used in the two-stage pretraining are presented in \cref{tab:hyper}. Upon completion of the training-relaxation-training paradigm, the foundation model DAO-G is fully pretrained and ready for finetuning on CSP datasets to specialize in the generation of stable crystal structures.

\subsection{Energy-Guided Sampling of DAO-G}\label{sec:energy_guidance}
Although we incorporate dataset relaxation during training to combat the instability arising from unstable data, this does not fully eliminate the issue. Therefore, we further address this challenge during sampling, drawing inspiration from the Energy-Based Models (EBMs).

We can learn the relationship between data distribution and the corresponding energy $\gE_0(\gM_0)$ from Boltzmann distribution~\cite{lifshitz1980_boltz}: $\gM_0 \sim \exp({-\beta\gE_0(\gM_0)}) / Z$, where $\beta$ is the temperature coefficient and $Z$ is the normalization constant. It indicates that data points with higher energy are less likely to occur. Importantly, the Boltzmann distribution exhibits a useful property: its log-likelihood is equivalent to the negative energy gradient with respect to data, which acts as a force field, driving updates towards lower energy configurations, and consequently, greater stability. Thus, the instability issue of unstable data incorporation is mitigated. This helpful conclusion motivates us to design an EBM to guide the sampling process.

The goal of those standard diffusion models is to learn the data distribution $q_0(\gM_0)$. Rather, we aim to train the energy-guided diffusion model to learn a modified distribution that is proportional to the original distribution weighted by the Boltzmann factor, namely, 
$p_0(\gM_0) \propto q_0(\gM_0) e^{-\beta\gE_0(\gM_0)}$. On this basis, CEP~\cite{lu2023cep} further proposes that: if the conditional distributions are equal, i.e., $p_{t0}(\gM_t | \gM_0)=q_{t0}(\gM_t | \gM_0)$, then the marginal distribution $p_t(\gM_t) \propto q_t(\gM_t) e^{-\beta\gE_t(\gM_t, t)}$, where $\gE_t(\gM_t)$ is the intermediate energy predicted by the pretrained DAO-P. The energy-guided distribution $p_t(\gM_t)$ adheres to equilibrium distribution better and its score is more interesting and informative, computed by:
\begin{equation}
\label{equ:new_score}
\underbrace{\nabla_{\gM_t}\log p_t(\gM_t)}_{\text{energy-guided score}} = \underbrace{\nabla_{\gM_t}\log  q_t(\gM_t)}_{\text{original score}}-\underbrace{\beta\nabla_{\gM_t}\gE_t(\gM_t)}_{\text{energy guidance}}.
\end{equation}
During the sampling phase of DAO-G, a pretrained DAO-P model is employed to provide energy-based guidance at each denoising step. Specifically, given the noisy intermediate structure $\mathcal{M}_t$ at timestep $t$ , DAO-P predicts its energy $f_\phi(\mathcal{M}_t, t)$. The gradient of this energy with respect to both lattice parameters and atomic coordinates, $\nabla_{\mathcal{M}_t} f_\phi(\mathcal{M}_t, t)$, is then computed and used as a guidance term. In parallel, DAO-G predicts the noise components $\epsilon_L$  (lattice) and $\epsilon_F$ (fractional coordinates) required for reconstructing $\mathcal{M}_{t-1}$. The final update combines the predicted noise with the energy gradient, steering the structure toward lower-energy configurations. In this way, DAO-P acts as an energy oracle, ensuring that the generative trajectory of DAO-G is biased toward physically plausible, energetically favorable structures.

\begin{figure}[t!]
\centering
\includegraphics[scale=0.6]{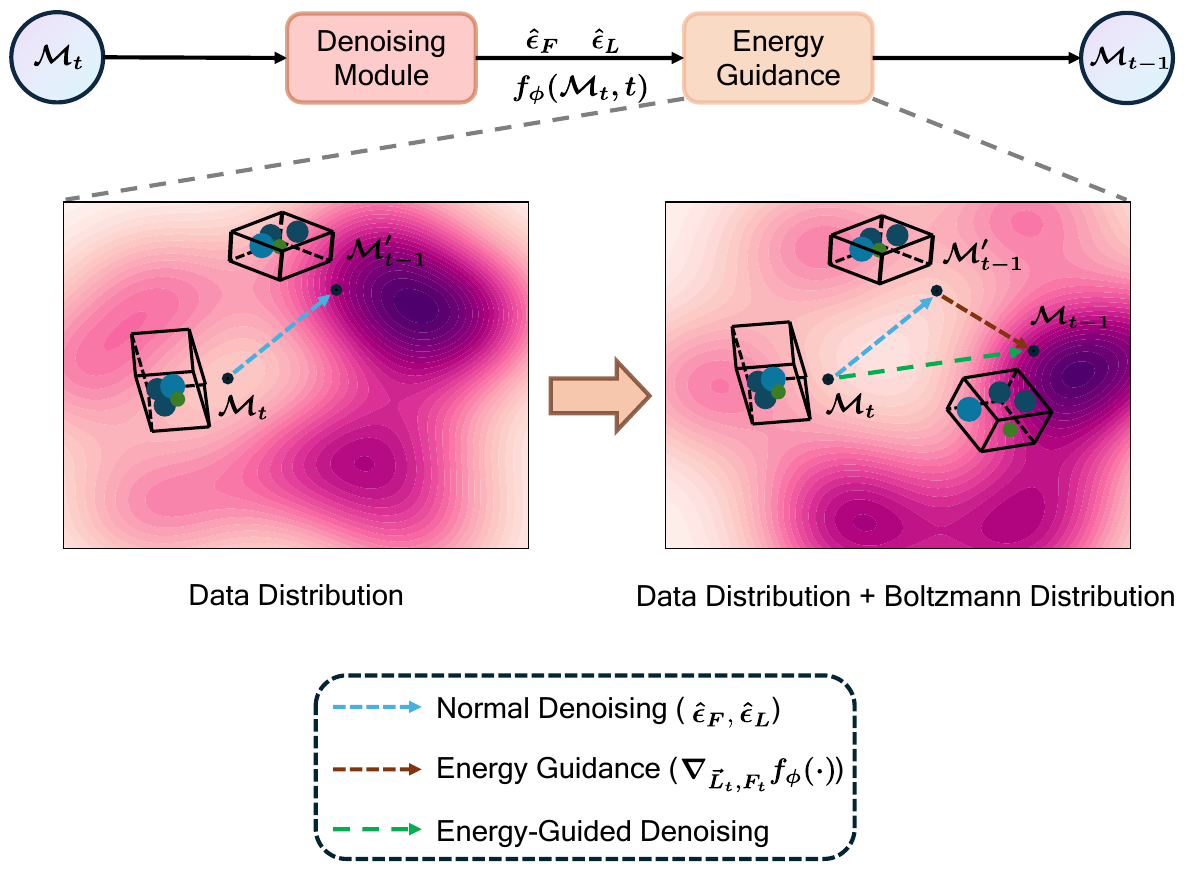}
\caption{The illustration of the energy-guided sampling process. The blue arrow represents standard denoising based on the data distribution, which, however, does not lie within stable regions. The brown arrow indicates the influence of energy guidance, steering the generation towards the equilibrium distribution. The resulting energy-guided denoising is depicted by the green arrow.}
\label{fig:energy_guidance}
\end{figure}

For implementation, we adapt the energy guidance strength using the property optimization method from DiffCSP~\cite{jiao2024diffcsp}. Two distributions are shown in \cref{fig:energy_guidance}: the data distribution (left) and the Boltzmann-augmented distribution (right). Standard sampling techniques, which prioritize high-probability regions within the data distribution, may not be suitable for capturing the Boltzmann distribution. By incorporating energy guidance, we can effectively sample from a distribution that respects both. \cref{alg:energy_guided_sampling} could provide a better understanding.

\begin{algorithm}[t!]
\caption{Energy-Guided Sampling of DAO-G}
\label{alg:energy_guided_sampling}
\begin{algorithmic}[1]
\State \textbf{Input:} Pretrained model $f_\phi$ for intermediate energy prediction, denoising model $f_\theta$ for noise prediction, input composition $\rmA$, denoising steps $T$, Langevin step size $\gamma$, guidance strength $s$, coefficient scheduler $\psi_L$ for lattice and $\psi_F$ for fractional coordinates.
\State \textbf{Output:} Generated Structure $\gM=(\rmA, \vec{\rmL}, \rmF)$.
\State Sample $\rmF_T\sim\gU(0,1), \vec{\rmL}_T\sim\gN(\vzero,\rmI).$ \Comment{Initialize $\rmF$ and $\vec{\rmL}$ from the predifined distributions.}
\For{step $t$ in $T,\, T-1,\, \dots\,,\,1$}
    \State Let $\gM_t=(\rmA, \vec{\rmL}_t, \rmF_t)$.
    \State{Sample $\vepsilon_\mL,\vepsilon_\rmF, \vepsilon'_\rmF\sim\gN(\vzero,\rmI)$.}
    \State{$\hat{\vepsilon}_\mL, \hat{\vepsilon}_\rmF \gets f_\theta(\gM_t, t)$.} \Comment{Predict the noises for lattice and coordinates.}
    \State {Calculate $\nabla_{\vec{\rmL}} E, \nabla_\rmF E$ for $E=f_\phi(\gM_t, t)$.} \Comment{The intermediate energy and gradients.}
    \State Acquire $[\alpha_t, \alpha_{t-1}]=\psi_L([t,t-1]), [\sigma_t, \sigma_{t-1}]=\psi_F([t, t-1]), \beta_t=1-\alpha_t,  \bar{\alpha}_t=\prod_{\tau=1}^t\alpha_\tau.$
    \State{$\vec{\rmL}_{t-1} \gets \frac{1}{\sqrt{\alpha_t}} (\vec{\rmL}_t - \frac{\beta_t} {\sqrt{1 -\bar{\alpha}_t}}\hat{\vepsilon}_\mL) - s\beta_t\cdot\frac{1-\bar{\alpha}_{t-1}}{1-\bar{\alpha}_t} \nabla_{\vec{\rmL}} E  + \sqrt{\beta_t\cdot\frac{1-\bar{\alpha}_{t-1}}{1-\bar{\alpha}_t}}\vepsilon_\mL$.} \Comment{Adding guidance.}
    \State{$\rmF_{t-\frac{1}{2}} \gets w(\rmF_t + (\sigma_t^2-\sigma_{t-1}^2)\hat{\vepsilon}_\rmF - s\frac{\sigma_{t-1}^2 (\sigma_t^2-\sigma_{t-1}^2)}{\sigma_t^2} \nabla_\rmF E + \frac{\sigma_{t-1}\sqrt{\sigma_t^2-\sigma_{t-1}^2}}{\sigma_t} \vepsilon_\rmF)$.} 
    \State Let $\gM_{t-\frac{1}{2}}=(\rmA, \vec{\rmL}_{t-1}, \rmF_{t-\frac{1}{2}})$. \Comment{Predictor-Corrector for coordinates.}
    \State{$\_, \hat{\vepsilon}_\rmF \gets f_\theta(\gM_{t-\frac{1}{2}}, t - 1)$.}
    \State{$d_t\gets \gamma\sigma_{t-1} / \sigma_1$.}
    \State{$\rmF_{t-1} \gets w(\rmF_{t-\frac{1}{2}} + d_t \hat{\vepsilon}_\rmF + \sqrt{2 d_t}\vepsilon'_\rmF)$.}
\EndFor
\State \textbf{Return} $\gM=(\rmA, \vec{\rmL}_0, \rmF_0)$. \Comment{Return the generated crystal.}
\end{algorithmic}
\end{algorithm}

\subsection{Mix-Supervised Pretraining for DAO-P}\label{sec:daop_pretraining}
In our work, DAO-P plays different roles depending on the task:
\begin{itemize}
    \item For the generation task (CSP in \cref{sec:exp_structure_generation}): DAO-P is used to support the training of DAO-G, providing dataset relaxation and energy-guided sampling. Since the CSP task requires generating structures purely from compositions, we must ensure that neither DAO-G nor DAO-P is exposed to the crystal structures of the downstream test set in advance. Therefore, both DAO-G and DAO-P are pretrained on the deduplicated CrysDB.
    \item For the property prediction tasks: DAO-P is used as a pretrained model and is finetuned on various datasets for property prediction. In this case, the model learns structure-to-property mappings. We have carefully considered the potential risk of data leakage. Although DAO-P is pretrained on CrysDB with labels of Ehull (extracted from MP and OQMD), our downstream tasks involve a different set of labels (e.g., JARVIS\_Ehull), which are not the same. Thus, pretraining DAO-P on the non-deduplicated CrysDB is acceptable, and using a larger dataset allows DAO-P to learn more robust structural representations.
\end{itemize}

Moreover, DAO-P provides crucial energy guidance during the sampling process of DAO-G. 
Inspired by the energy-guided diffusion~\cite{lu2023cep}, our target distribution of crystal $\gM$ is defined as 
$p(\gM) \propto q(\gM)e^{-\beta \gE(\gM)}$, where $q(\gM)$ is the data distribution and $e^{-\beta \gE(\gM)}$ is the Boltzmann distribution with
$\gE(\gM)$ being an energy function. By adding the Boltzmann term, we encourage the generated structures to have lower energies and thus be more stable. 
During the forward diffusion process, we denote the intermediate energy function as $\gE_t(\gM_t)$ at each timestep $t$, leading to intermediate distributions 
$p_t(\gM_t) \propto q_t(\gM_t)e^{-\gE_t(\gM_t)}$.  
Therefore, in order to ensure that the reverse process produces correct samples from $p(\gM)$, it is essential to model not only the final energy $\gE_0(\gM_0)$ but also its intermediate counterparts $\gE_t(\gM_t)$. 
This motivates the introduction of an intermediate energy loss. Predicting intermediate energy is non-trivial because $\gE_t(\gM_t)$ is defined as a log-expectation~\cite{lu2023cep}: 
\begin{equation}
\label{equ:inter_energy_expression}
    \gE_t(\gM_t) = -\log \mathbb{\gE}_{q_0(\gM_0|\gM_t)}\!\left[e^{-\beta \gE(\gM_0)}\right],
\end{equation}
which cannot be computed in closed form except at $t=0$. 
This makes direct supervision unavailable. Previous approaches attempted to solve it by employing techniques such as MSE supervision~\cite{janner2022diffusion_planning,bao2022EEGSDE}, posterior sampling~\cite{ho2022vdm} and contrastive prediction~\cite{lu2023cep}. Nevertheless, these methods suffer from either inaccurate prediction or computational instability~\cite{wang_confdiff,lu2023cep}.

For this functionality, we pretrain DAO-P using a hybrid loss function that combines self-supervised diffusion losses (\cref{equ:loss_F,equ:loss_L}, as in DAO-G pretraining) and a supervised intermediate energy prediction loss (\cref{equ:IEP_loss}). Given the energy predicted by Crysformer's energy head (\cref{sec:crysformer}), we denote the collective parameters as $\phi$ and the resulting predicted energy as $f_\phi(\gM_t, t)$. The energy loss utilizes the exponential loss function with martingale policy rather than the direct MSE prediction:
\begin{equation}
    \label{equ:IEP_loss}
    \phi^*=\argmin_{\phi}\E_{q_{0t}(\gM_0,\gM_t)}\left[\|e^{-f_\phi (\gM_t,t)}-e^{-\beta\gE_0(\gM_0)} \|_2^2 \right].
\end{equation}
This formulation bypasses the intractable log-expectation, provably yields the exact gradient of $\gE_t(\gM_t)$. We presented the proof in \cref{sec:loss_design}.
In this way, our design provides a principled and effective solution where existing approaches fall short.

For downstream energy prediction, although only the energy prediction head of DAO-P is directly utilized, the inclusion of lattice and coordinate denoising heads is still beneficial. These denoising tasks act as self-supervised signals that encourage the model to learn richer structural representations. Prior work has demonstrated the effectiveness of such diffusion-based pretraining in improving representation learning for crystal property prediction~\cite{song2024diffusion}, and similar approaches have proven useful in other domains such as molecular property prediction~\cite{zaidi2022pre} and semantic segmentation~\cite{brempong2022denoising}. In our case, the incorporation of noise prediction heads enhances DAO-P's generalization capacity for energy predictions while simultaneously enabling effective finetuning for additional property prediction tasks.

\subsection{Superconductor Dataset Augmentation}\label{sec:supercon3d_aug}
We begin with a brief overview of the SuperCon dataset~\cite{supercon}. This dataset contains approximately 33,000 superconductors, providing only their chemical formulas and corresponding critical temperatures $T_c$. A recent curation effort~\cite{chen2025supercon3d} identified structures for some of these superconductors within the ICSD database~\cite{bergerhoff1983icsd}, creating a structural subset called SuperCon3D. Our focus on ordered superconductors yields a dataset of 1017 structural entries (SC-s) and 748 entries without associated structures (SC-ns).

\begin{figure}[t!]
\centering
\includegraphics[scale=0.95]{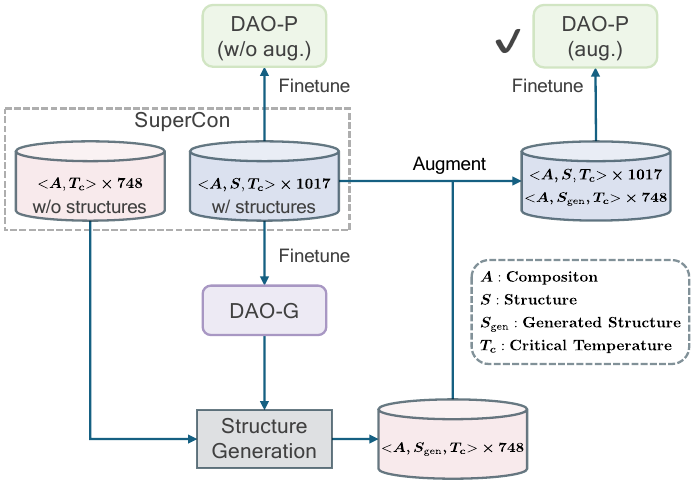}
\caption{The finetuning process of DAO-G and DAO-P on the curated SuperCon dataset~\cite{supercon,chen2025supercon3d}, in which 3D structures are known for only a subset of materials. ``aug.'' denotes the data augmentation.}
\label{fig:supercon_flowchart}
\end{figure}

To evaluate the ability of DAO-P to predict superconductors $T_c$, we perform 5-fold cross-validation on SC-s and the model is referred to as DAO-P (w/o aug.). It is worthwhile to investigate whether SC-ns can contribute to enhanced performance. We explore this possibility by employing  DAO-G and DAO-P in a cooperative way. Specifically, we finetune the pretrained DAO-G on the full SC-s, adapting its generative capabilities to superconductor domain. This allows us to generate structures for the SC-ns entries, enriching the training data for each fold. The model finetuned on the augmented dataset is denoted as DAO-P (aug.). The whole process is depicted in \cref{fig:supercon_flowchart}.

\subsection{Evaluation Metrics}\label{sec:metrics}
We assess structure generation performance with two metrics: Match Rate (MR), which measures the percentage of successful matches between generated and ground truth structures, and Root Mean Square Error (RMSE), which quantifies the deviation in atomic coordinates. For the calculation of MR and RMSE, we use the StructureMatcher class from Pymatgen library~\cite{ong2013_pymatgen}, with thresholds stol=0.5, angle\_tol=10, ltol=0.3. The mathematical formulations are expressed as follows:
\begin{align}
\text{MR} &= \frac{|\mathcal{I}|}{M}, \ \ \ \mathcal{I} = \left\{ i \in \{1, \dots, M\} \mid \text{StructureMatcher}(G_i, T_i) = \text{True} \right\}, \\
\text{RMSE} &= \frac{1}{|\mathcal{I}|} \sum_{i \in \mathcal{I}} \left( \frac{1}{\sqrt[3]{V_i / N_i}} \sqrt{ \sum_{j=1}^{N_i} \| \mathbf{r}_{i,j} - \mathbf{r}'_{i,j} \|^2 } \right),
\end{align}
where, $M$ is the total number of generated samples, $G_i$ is the $i$-th generated structure, $T_i$ is the corresponding $i$-th ground truth structure, and $\mathcal{I}$ denotes the subset of successfully matched data indices; $V_i$ is the volume of the $i$-th crystal, $N_i$ is its atom number, $\mathbf{r}_{i,j}$ and $\mathbf{r}_{i,j}' $ are the Cartesian coordinates of the $j$-th matched atom in $G_i$ and Structure $T_i$, respectively, and $\| \cdot \|$ denotes the Euclidean distance. The StructureMatcher operates within a tolerance framework governed by three key parameters: stol (tolerance for maximum distance between atomic sites), ltol (tolerance for lattice vector lengths), and angle\_tol (tolerance for lattice angles, in degrees). It attempts to seek a one-to-one atomic correspondence that satisfies all tolerance thresholds. If such a mapping is found, it returns True; otherwise, it returns False. Note that before computing the interatomic distances for RMSE, the structures $G_i$ and $T_i$ are first optimally aligned to achieve the closest structural match (automatically handled by StructureMatcher, requiring no manual intervention). Regarding 20-shot generation, for each target structure $T_i$, we generate 20 candidate structures. Each generated structure is first compared with $T_i$ using the StructureMatcher. Among the subset of candidates that are successfully matched, we select the one with the smallest structural error as $G_i$ (i.e., $
\argmin_{k \in \{1, \dots, 20\}} \left( \frac{1}{\sqrt[3]{V_i / N_i}} \sqrt{ \sum_{j=1}^{N_i} \| \mathbf{r}_{i,j} - \mathbf{r}'_{k,j} \|^2 } \right)$). After determining 
$G_i$ for all target structures in this manner, the Match Rate and RMSE are computed according to the above formulas.

Property prediction is evaluated using Mean Absolute Error (MAE) between predicted and labeled properties. While for superconductor property prediction, we use MAE calculated on the logarithm of the $T_c$ values. The metric of MAE can be formulated as:
\begin{equation}
    \text{MAE} = \frac{1}{N_p} \sum_{i=1}^{N_p} |y_i - y'_i|,
\end{equation}
where, $N_p$ is the total number of data points, $y_i$ is the reference value (Reverse standardization), $y'_i$ is the predicted value (Reverse standardization), $|\cdot|$ denotes the absolute value.

\section*{Data Availability}
The raw datasets for pretraining and downstream tasks are available publicly at the following links:
\begin{itemize}
    \item Materials Project~\cite{jain2013_mp_dataset}: \href{https://next-gen.materialsproject.org}{https://next-gen.materialsproject.org}.
    \item OQMD~\cite{kirklin2015_oqmd}: \href{https://oqmd.org}{https://oqmd.org}.
    \item MatBench~\cite{dunn2020_matbench}: 
    \href{https://matbench.materialsproject.org}{https://matbench.materialsproject.org}.
    \item JARVIS-DFT 3D~\cite{choudhary2020_jarvis3d}: 
    \href{https://figshare.com/articles/dataset/jdft_3d-7-7-2018_json/6815699}{https://figshare.com/articles/dataset/jdft\_3d-7-7-2018\_json/6815699}.
\end{itemize}
The processed datasets are collected and available at \href{https://doi.org/10.6084/m9.figshare.31440697}{https://doi.org/10.6084/m9.figshare.31440697}. Source data are provided with this paper.

\section*{Code Availability}

The source code developed for this study, together with a detailed README file, is publicly available at the \href{https://github.com/ManlioWu/DAO}{https://github.com/ManlioWu/DAO} with DOI: \href{https://doi.org/10.5281/zenodo.18824014}{https://doi.org/10.5281/zenodo.18824014}.

\putbib[reference]

\section*{Acknowledgements}
We would like to express our sincere gratitude to Mr. Qi Li, Prof. Shifeng Jin from the Chinese Academy of Sciences, Dr. Mengshi Wang from Huawei, and Mr. Mingze Li, Mr. Songyou Li, Mr. Anyi Li, Mr. Jiacheng Cen, Mr. Yuelin Zhang from Renmin University of China for their invaluable discussions and insightful suggestions. Their contributions have greatly enhanced the quality of this work. The research is financially supported by the National Natural Science Foundation of China (No. 62376276), the Beijing Nova Program (No. 20230484278), the Fundamental Research Funds for the Central Universities, the Research Funds of Renmin University of China (23XNKJ19), and the Public Computing Cloud at Renmin University of China.

\section*{Author Contributions Statement}
L.W. and W.H. drafted the manuscript; L.W. and R.J. developed the code and conducted the experiments under the supervision of W.H.; L.W. organized the experimental results; W.H., J.H., L.L., Y.Z., H.S., Y.L., F.S., Y.R., and J.W. provided technical support; W.H., Y.R., and J.W.  supervised the research; W.H. led the project; all authors reviewed and approved the final version of the manuscript.

\section*{Competing Interests Statement}
The authors declare no competing interests.

\label{LastMainPage}
\end{bibunit}

\clearpage
\appendix
\renewcommand{\refname}{Supplementary References}
\begin{bibunit}[unsrt]
\refstepcounter{page}
\label{FirstAppendixPage}

\newpage
\renewcommand{\thesection}{\arabic{section}} %

\setlength{\cftsubsecindent}{2em}
\setlength{\cftsubsubsecindent}{4em}

\setlength{\cftbeforesecskip}{7pt}
\setlength{\cftbeforesubsecskip}{5pt} 
\setlength{\cftbeforesubsubsecskip}{3pt}

\renewcommand{\cftsecleader}{\cftdotfill{\cftdotsep}\hspace{0.5em}}
\renewcommand{\cftsubsecleader}{\cftdotfill{\cftdotsep}\hspace{0.5em}} 
\renewcommand{\cftsubsubsecleader}{\cftdotfill{\cftdotsep}\hspace{0.5em}}

\renewcommand{\cftsecpagefont}{\bfseries\color{magenta}} 
\renewcommand{\cftsubsecpagefont}{\color{magenta}} 
\renewcommand{\cftsubsubsecpagefont}{\color{magenta}} 

\crefalias{table}{app:table}
\crefalias{figure}{app:figure}
\crefalias{algorithm}{app:algorithm}


\title{Supplementary Information: Siamese Foundation Models for Crystal Structure Prediction}
\renewcommand{\abstracttitle}{}
\renewcommand{\keywordname}{}
\begin{abstract}
\end{abstract}
\keywords{}
\settitle

\vspace{-1in}

\startlist{toc}
\printlist{toc}{}{\section*{Contents}}

\setcounter{page}{1}
\setcounter{figure}{0}
\setcounter{table}{0}

\renewcommand{\tablename}{Supplementary Table}
\renewcommand{\figurename}{Supplementary Figure}

\renewcommand\thesection{\Alph{section}}
\renewcommand\thesubsection{\thesection.\arabic{subsection}}
\renewcommand\thesubsubsection{\thesubsection.\arabic{subsubsection}}

\graphicspath{ {./Figure_supp/} }
\newcommand{\eref}[1]{(\ref{#1})}

\defaultbibliographystyle{unsrt}

\clearpage
\section{Preliminaries}\label{sec:preliminary}

\subsection{Diffusion Models}\label{sec:intro_of_diffusion} 
The diffusion process unfolds in two stages: a forward pass that adds noise to the data and a backward pass focused on denoising. During the forward process, noise is progressively added to the original data sample until it conforms to a predefined distribution at timestep $T$, commonly chosen to be a standard normal distribution. Denoising (the backward process) often requires the model to predict the noise introduced at each timestep, as in DDPM~\cite{supp:ho2020ddpm}. Alternatively, methods like Score-Matching~\cite{supp:song2021maximum} aim to predict the gradient of the log-likelihood, also known as the score $s_\theta$. When utilizing diffusion models for the generation of geometric objects, particularly those in three dimensions, it is imperative to incorporate knowledge of the inherent symmetry present in the data to avoid generating unrealistic or physically implausible structures. For instance, several studies~\cite{supp:hoogeboom2022edm,supp:huang2023mdm} have incorporated $\mathrm{E(3)}$ equivariance into their diffusion model architectures for molecular generation. In the same vein,  our diffusion-based model for crystal structure pretraining needs to account for this fundamental constraint, which is discussed in the following.

\subsection{Equivariance and Invariance} \label{sec:equivariance_and_invariance}
We investigate the equivariance of geometric models in the group $\mathfrak{G}$, specifically analyzing how the transformation of the input structure affects the output~\cite{supp:cen2024high,supp:satorras2021n,supp:cen2025universally}. A function $f:\gX \to \gY$ exhibits equivariance if it satisfies $\rho_{\gY}(\mathfrak g)\cdot f(x)=f(\rho_{\mathcal X}(\mathfrak g)\cdot x), \forall \mathfrak g \in \mathfrak {G}$, where $\rho_{\gX}(\mathfrak g)$ and $\rho_{\mathcal 
 Y}(\mathfrak g)$ denote group representations of group element $\mathfrak g$ in input and output spaces, respectively. When $\rho_{\gY}(\mathfrak g)$ is a trivial representation (\emph{i.e.}, always identity transformation), the function $f$ is invariant to the transformation; otherwise, it is termed equivariant. Hence, invariance can be considered a special case of equivariance.  Among those applications in biochemistry, it is a common practice to consider Euclidean transformation (\emph{i.e.}, $\mathrm{E}(3)$ Group), including rotation, translation and reflection. Specially owing to the periodic nature of crystalline structures, it is necessary to take periodical translation invariance into account additionally. A thorough discussion of the symmetry considerations is provided as follows:

\paragraph{$\mathrm{O}(3)$-invariance.} For any rotation/reflection matrix $\rmO\in\mathrm{O}(3)$ imposed on the lattice matrix $\vec{\rmL}$, there is $p(\rmO\vec{\rmL}, \rmF | \rmA)=p(\vec{\rmL}, \rmF | \rmA)$. The distribution remains unchanged no matter how the lattice is transformed under $\mathrm{O}(3)$ group.

\paragraph{Periodical translation invariance.} Upon translating all atoms by $\rvt\in\sR^{3\times 1}$, the resulting fractional coordinates are expressed as $\{\rmF+\rvt\bm{1}^\top\}$, with $\{\cdot\}$ representing the fractional part extraction. Then the equation $p(\vec{\rmL}, \{\rmF+\rvt\bm{1}^\top\} | \rmA)=p(\vec{\rmL}, \rmF | \rmA)$ follows, illustrating the fact that translating the atom coordinates does not affect the distribution.
\newline

\noindent For simplicity, we collectively refer to the $\mathrm{O}(3)$-invariance and periodical translation invariance as OP-invariance.

\section{Theoretical Analysis}\label{sec:theory_analysis}
In this section, we provide rigorous mathematical proofs and detailed derivations for the key theoretical results presented in the main text.

\subsection{Invariance of the learned Node Features}
\label{sec:model_invariance_proof}
In this part, we provide the details why $\rmH^{(S)}$ is OP invariant. From Appendix A of DiffCSP~\cite{supp:jiao2024diffcsp}, we can learn that
\begin{equation}
    \bm\varphi_{\texttt{FT}}\left(\{\rvf_{i}+\rvt\} - \{\rvf_{j}+\rvt\}\right)=\bm\varphi_{\texttt{FT}}\left(\rvf_{i} -\rvf_{j}\right),
\end{equation} 
where, $\rvt\in\sR^{3\times 1}$ is the translation. Then, we prove that the pre-calculated edge feature $\rve_{ij}$ exhibits OP-invariance. Given any rotation/reflection matrix $\rmO\in\mathrm{O}(3)$ imposed on the lattice matrix $\vec{\rmL}$, we have: 
\begin{equation}
\begin{aligned}
\rve_{ij}^{\texttt{transformed}}&=\texttt{Normalization}\left((\rmO\vec{\rmL})^\top (\rmO\vec{\rmL}) \ \| \bm\varphi_{\texttt{FT}}\left(\{\rvf_{i}+\rvt\} - \{\rvf_{j}+\rvt\}\right)\right),\\
&=\texttt{Normalization}\left(\vec{\rmL}^\top(\rmO^\top\rmO)\vec{\rmL} \ \| \ \bm\varphi_{\texttt{FT}}\left(\rvf_{i} -\rvf_{j}\right)\right) \\
&=\texttt{Normalization}\left(\vec{\rmL}^\top\vec{\rmL} \ \| \ \bm\varphi_{\texttt{FT}}\left(\rvf_{i} - \rvf_{j} \right)\right)\\
&=\rve_{ij}.
\end{aligned}
\end{equation}
Moreover, the node feature $\rvh_i^{(0)}$ is OP-invariant, because it is initialized by CGCNN embedding, depending solely on the atom type $\va_i$. Therefore, the inputs (i.e. $\bm{E}, \rmH^{(0)}$) to our noise prediction model are all OP-invariant. Since the components of our Transformer block, including multi-head attention and feedforward layers, operate exclusively on OP-invariant features, the final output $\rvh_i^{(S)}$ is also OP-invariant.

\subsection{Equivariance of the Noise Output}
\label{sec:noise_equivariance_proof}
For fractional coordinates, we compute the score by $\hat{\bm{\epsilon}}_F(\gM_t, t)=\bm\varphi_F(\rmH^{(S)}_t)$. As we have proved that $\rmH^{(S)}_t$ is OP-invariant in \cref{sec:model_invariance_proof}, it is obvious that the predicted $\hat{\bm{\epsilon}}_F(\gM_t, t)$ also meets OP-invariance. In terms of the lattice noise, we just need to consider $\mathrm{O(3)}$ transformation. For $\hat{\bm{\epsilon}}_L(\gM_t, t)=\vec{\rmL}\bm\varphi_L(\bar{\rvh}_t)$, if we impose an $\mathrm{O(3)}$ transformation $\rmO$ on the crystal, $\bar{\rvh}_t$ remains unchanged and the lattice representation becomes $\rmO\vec{\rmL}$. We derive the transformed score:
\begin{equation}
    \begin{aligned}
        \hat{\bm{\epsilon}}_L^{\texttt{transformed}}(\gM_t, t)&=(\rmO\vec{\rmL})\bm\varphi_L(\bar{\rvh}_t) \\
        &=\rmO\left(\vec{\rmL}\bm\varphi_L(\bar{\rvh}_t)\right)\\
        &=\rmO\hat{\bm{\epsilon}}_L(\gM_t, t).
    \end{aligned}
\end{equation}
It indicates that the noise $\hat{\bm{\epsilon}}_L(\gM_t, t)$ is also transformed given the transformation on the input crystal. That is, the predicted score for lattice is $\mathrm{O(3)}$ equivariant. Since the crystal representation $\bar{\rvh}_t$ is OP-invariant, the energy output $\bm\varphi_E(\bar{\rvh}_t)$ maintains OP-invariance as well.

\subsection{Derivation of the Intermediate Energy Prediction Loss}
\label{sec:loss_design}

From CEP~\cite{supp:lu2023cep}, we know that the intermediate energy  $\gE_t(\gM_t)=-\log \mathbb
{E}_{q_{0t}(\gM_0|\gM_t)} [e^{-\beta\gE_0(\gM_0)}]$, that is $e^{-\gE_t(\gM_t)}=\mathbb
{E}_{q_{0t}(\gM_0 | \gM_t)} [e^{-\beta\gE_0(\gM_0)}]$. To model $\gE_t(\gM_t)$ for any $t > 0$, we design a network $f_\phi(\gM_t, t)$ to approximate it by the loss function shown in \cref{equ:IEP_loss}. Here we prove this loss function can lead to the objective.

\begin{equation}
\begin{aligned}
\gL_\texttt{IEP}(\phi)&=\E_{q_{0t}(\gM_0,\gM_t)}\left[\|e^{-f_\phi (\gM_t,t)}-e^{-\beta\gE_0(\gM_0)} \|_2^2 \right] \\
 &=\E_{q_{0t}(\gM_0,\gM_t)}\left[e^{-2f_\phi (\gM_t,t)}-2e^{-f_\phi (\gM_t,t)-\beta\gE_0(\gM_0)}+e^{-2\beta\gE_0(\gM_0)} \right] \\
 &=\E_{q_{0t}(\gM_0,\gM_t)}\left[e^{-2f_\phi (\gM_t,t)}\right]-2\E_{q_{0t}(\gM_0,\gM_t)}\left[e^{-f_\phi (\gM_t,t)-\beta\gE_0(\gM_0)} \right]+C_1 \\
 &=\E_{q_{t}(\gM_t)}\left[e^{-2f_\phi (\gM_t,t)}\right]-2\E_{q_{t}(\gM_t)}\E_{q_{0t}(\gM_0|\gM_t)}\left[e^{-f_\phi (\gM_t,t)-\beta\gE_0(\gM_0)} \right]+C_1 \\
 &=\E_{q_{t}(\gM_t)}\left[  e^{-2f_\phi (\gM_t,t)} -2\E_{q_{0t}(\gM_0|\gM_t)}\left[e^{-f_\phi (\gM_t,t)-\beta\gE_0(\gM_0)}\right ] \right]+C_1 \\
 &=\E_{q_{t}(\gM_t)}\left[  e^{-2f_\phi (\gM_t,t)} -2e^{-f_\phi (\gM_t,t)}\E_{q_{0t}(\gM_0|\gM_t)}\left[e^{-\beta\gE_0(\gM_0)}\right ] \right]+C_1,
\end{aligned}
\end{equation}
where $C_1=\E_{q_{0t}(\gM_0,\gM_t)}\left[  e^{-2\beta\gE_0(\gM_0)} \right]$ is a constant independent of the parameter $\phi$, and $\beta$ is the temperature. The optimal minimum of the loss function $\gL_\texttt{IEP}$ is attained if and only if the condition $e^{-f_\phi(\gM_t, t)}=\E_{q_{0t}(\gM_0 | \gM_t)} [e^{-\beta\gE_0(\gM_0)}]$ is satisfied, indicating that $f_\phi(\gM_t, t)$ serves as an approximation of $\gE_t(\gM_t)$. In the pretraining experiments, we set $\beta=1$.

\section{Downstream Datasets Introduction}
\paragraph{Matbench~\cite{supp:dunn2020_matbench}.} Matbench is a prevailing materials benchmark, tailored for property prediction. It is often used to evaluate various machine learning algorithms. We select JDFT2D~\cite{supp:choudhary2017_jdft2d} to predict Exfoliation Energy, Dielectric~\cite{supp:petousis2017_diel} to predict Refractive Index and KVRH~\cite{supp:de2015charting_kvrh} to predict the average Bulk Moduli. The experimental results are averaged by five folds, which are split by the benchmark in advance.

\paragraph{Materials Project~\cite{supp:jain2013_mp_dataset}.}  The Materials Project is a platform that offers access to computed materials data and advanced analysis tools. We follow the datasets used in MEGNet~\cite{supp:chen2019megnet}, to predict three properties: Shear Modulus, Bulk Modulus and Bandgap. For generation task, MP-20 and MPTS-52 are both sourced from MP. They are curated to contain materials with the atom numbers no more than 20 and 52, respectively.

\paragraph{JARVIS-3D~\cite{supp:choudhary2020_jarvis3d}.} We use Jarvis-Tools~\cite{supp:choudhary2024_jarvis_tools} to download the dft3d dataset. The version is aligned with Matformer~\cite{supp:yan2022matformer}. We choose properties of Ehull and Bandgap (MBJ), which are two challenging tasks.
\newline

\noindent Dataset size and the corresponding hyperparameters for each one are introduced in~\cref{sec:impl_details}.

\begin{table}[t!]
\centering
\caption{Hyperparameters at pretraining and finetuning stages. CrysDB denotes our Pretraining Dataset. DAO-P (pred.) is intended for property prediction, whereas DAO-P (gen.) is used for structure prediction. ``dedup'' denotes the deduplicated dataset.}
\label{tab:hyper}
\renewcommand{\arraystretch}{1.5}
\resizebox{\linewidth}{!}{
    \begin{tabular}{ccccccc} 
        \toprule
        Task & Dataset & \#Samples & Learning Rate & Weight Decay & Batch Size $\times$ GPUs & Epoch \\ 
        \hline
        Pretraining of DAO-P (pred.) & CrysDB &942,884 & 2e-4 & 0 & 1024$\times$2 & 800 \\ 
        Pretraining of DAO-P (gen.) & CrysDB (dedup) & 919,258 & 2e-4 & 0 & 1024$\times$2 & 800 \\ 
        Pretraining of DAO-G (Stage \Rst) & CrysDB (dedup) &919,258 & 3e-4 & 0 & 1024$\times$3 & 800 \\ 
        Pretraining of DAO-G (Stage \Rnd) & CrysDB (dedup, relaxed) &919,258 & 1e-4 & 0 & 1024$\times$3 & 500 \\ 
        \hline
        \multirow{9}{0.215\linewidth}{\hspace{0pt}Property Prediction} & JDFT2D & 636 & 8e-5 & 5e-5 & 128$\times$1 & 300 \\
         & DIELECTRIC & 4,764 & 5e-5 & 5e-5 & 24$\times$2 & 300 \\
         & KVRH &10,987 & 8e-5 & 1e-5 & 128$\times$1 & 200 \\
         & Jarvis\_gap &18,171 & 1e-4 & 2e-4 & 256$\times$2 & 300 \\
         & Jarvis\_Ehull &55,370 & 7e-4 & 1e-4 & 256$\times$3 & 500 \\
         & Mp\_Shear &5,449 & 1e-4 & 2e-5 & 256$\times$2 & 500 \\
         & Mp\_Bulk &5,450 & 3e-4 & 0 & 256$\times$2 & 500 \\
         & MP\_gap &69,239 & 2e-5 & 0 & 64$\times$3 & 300 \\
        \hline
        \multirow{2}{0.215\linewidth}{\hspace{0pt}Structure Generation} & MP-20 &45,231 & 2e-5 & 1e-5 & 400$\times$8 & 1000 \\
         & MPTS-52 &40,476 & 2e-5 & 1e-5 & 80$\times$8 & 1000 \\
        \bottomrule
    \end{tabular}
}
\end{table}

\begin{table}[t!]
    \centering
    \caption{Task and Resource Allocation}
    \label{tab:gpu_hour}
    \resizebox{0.5\linewidth}{!}{
    \begin{tabular}{lcccc}
        \toprule
        \textbf{Task} & \textbf{Model} & \textbf{GPU Type} & \textbf{GPU Days} \\
        \midrule
        \multirow{4}{*}{Pretrain} 
            & DAO-G (Stage \Rst) & A100 (80G) & 3.85 \\
            & DAO-G (Stage \Rnd) & A100 (80G)  & 3.12 \\
            & DAO-P  (dedup) & A100 (80G) & 3.83 \\
            & DAO-P & A100 (80G) & 4.52 \\
        \midrule
        \multirow{2}{*}{Finetune} 
            & DAO-G (mp-20) & 4090 (24G) & 0.34 \\
            & DAO-G (mpts-52)& 4090 (24G) & 0.85 \\
        \bottomrule
    \end{tabular}
    }
\end{table}

\section{Implementation Details}\label{sec:impl_details}
In this section, we present the used hyperparameters of our models and pretrained CSP baselines, and the configurations of finetuning.

\subsection{Hyperparameters of DAO}
Our pretrained model is a 12-layer Crysformer architecture with a hidden dimension of 384, 8 attention heads, and a SiLU activation function, resulting in a model size of 25M parameters. We adopt the Adam optimizer for training the network and utilize a cosine learning rate scheduler with linear warmup for learning rate adjustment, consistent with current trends in large language models (LLMs) training. 
\cref{tab:hyper} outlines specific hyperparameters during pretraining and finetuning stages, and \cref{tab:gpu_hour} lists the resource consumption of pretraining and finetuning (on CSP tasks) process.

\subsection{Hyperparameters of Pretrained CSP Baselines}
\label{sec:csp_baseline_hparams}
In our experiments, we carefully tuned the hyperparameters for each pretrained CSP baseline on the MP-20 and MPTS-52 datasets. For MatterGen, we employed 2 blocks and used a learning rate of $1 \times 10^{-4}$ on both datasets; For DiffCSP, the learning rate was set to $5 \times 10^{-5}$ on MP-20 and $1 \times 10^{-4}$ on MPTS-52, with a weight decay of $1 \times 10^{-5}$ applied consistently across both datasets. The larger variant, DiffCSP-large, used a learning rate of $5 \times 10^{-5}$ on MP-20 and $2 \times 10^{-5}$ on MPTS-52, also with a weight decay of $1 \times 10^{-5}$. For FlowMM, we configured the model with a hidden dimension of 512 and 11 message-passing layers, using a uniform learning rate of $1 \times 10^{-4}$ on both datasets. Finally, FlowMM-Crysformer was implemented with a hidden dimension of 480 and 5 layers, also trained with a learning rate of $1 \times 10^{-4}$ on MP-20 and MPTS-52. All unspecified parameters were kept at their default values.

\subsection{Configurations of Finetuning}
For crystal property prediction downstream tasks, a two-layer MLP (i.e., prediction head) is integrated with the pretrained model, allowing for simultaneous finetuning of the entire model. Separate prediction heads are used for each dataset and when training we normalize the property labels to follow a standard normal distribution for better numerical stability. 

While for CSP tasks, we finetune the model using the same loss functions as the pretraining. After finetuning, we generate structures with the test set and evaluate them against the ground truth by calculating the MR and RMSE.

\section{Extended Experimental Results}
\subsection{Deeper Analysis of Structure Generation}\label{sec:detailed_comparison_of_DAO-G}
In this section, we present further results on the CSP task, including ablation studies of DAO-G and its 20-shot sampling performance.

\subsubsection{Ablations on Two-Stage Pretraining}\label{sec:dao_ablation}
The finetuning results for various DAO configurations on the MP-20 and MPTS-52 datasets are presented in \cref{tab:DAO-G_comprison}. It can be observed that:

\begin{table*}[t!]
  \centering
  \setlength{\tabcolsep}{10pt}
  \caption{Experimental results (averaged on three runs) for CSP task across different DAO-G configurations. Here, ``Stage \Rst'' denotes first-stage-only pretraining, ``stable'' represents pretraining using the 306,830 stable crystals, ``only relax'' means using the relaxed data only, and ``w/ guidance'' indicates using energy-guided sampling.}
  \label{tab:DAO-G_comprison}
  \vspace{0.03in}
  \resizebox{0.98\textwidth}{!}{
    \begin{tabular}{lccccc}
    \toprule
    \multirow{2}[2]{*}{} & \textbf{\# of} & \multicolumn{2}{c}{\textbf{MP-20}} & \multicolumn{2}{c}{\textbf{MPTS-52}} \\
          &  \textbf{samples}   & \textbf{Match Rate (\%) $\uparrow$} & \textbf{RMSE $\downarrow$} & \textbf{Match Rate (\%) $\uparrow$} & \textbf{RMSE $\downarrow$} \\
    \midrule
    DAO-G (Stage \Rst, stable)  &1 & 65.63 ± 0.07	& 0.0434 ± 0.0003	& 31.17 ± 0.11	& 0.0790 ± 0.0015 \\ \midrule
    DAO-G (Stage \Rst)  &1 & 65.60 ± 0.22	& 0.0411 ± 0.0005 & 32.52 ± 0.18	& 0.0731 ± 0.0008 \\ \midrule
    DAO-G (Stages \Rst+\Rnd)  &1 & \textbf{65.97 ± 0.22}	& \underline{0.0401 ± 0.0013} & \underline{32.59 ± 0.10}	& 0.0695 ± 0.0013 \\ \midrule
    DAO-G (Stages \Rst+\Rnd, w/ guidance)  &1 & 65.65 ± 0.18	& 0.0406 ± 0.0014 & \textbf{32.78 ± 0.06}	&\underline{0.0688 ± 0.0023} \\ \midrule
    DAO-G (Stage \Rst, only relax) & 1 &58.37 ± 0.25 &0.0605 ± 0.0019 & 23.78 ± 0.16 & 0.0964 ± 0.0004 \\ \midrule 
    DAO-G (Stage \Rst+\Rnd\,(stable+relax)) & 1 & 65.29 ± 0.14 & \textbf{0.0387 ± 0.0002} & 32.16 ± 0.03 & \textbf{0.0685 ± 0.0011} \\ \midrule
    DAO-P (gen.) & 1 & 63.01 ± 0.02 & 0.0480 ± 0.0007 & 29.73 ± 0.20 & 0.0819 ± 0.0007 \\
    \bottomrule
    \end{tabular}%
}
\end{table*}

\begin{itemize}
    \item Stage \Rst\, vs. Stage \Rst\,(stable). We compare two settings for Stage \Rst\, pretraining: (i) using the full deduplicated dataset (including unstable structures) and (ii) using only the stable subset. The results show that, except for the Match Rate metric on MP-20 where the two settings are comparable, the full-dataset model consistently outperforms the stable-only model across other benchmarks. This confirms that incorporating unstable structures is indeed reasonable.
    \item DAO-G (Stage \Rst) vs. DAO-P (gen.). Since DAO-P is trained with both structure generation and energy prediction losses, in principle it is capable of crystal structure generation. However, DAO-P is optimized under a composite objective—balancing both energy regression and denoising for structure generation—within the same model capacity as DAO-G. This trade-off inevitably weakens its ability in structure generation compared to a dedicated model. To verify this, we directly evaluated DAO-P on the CSP task. The results show that DAO-P underperforms DAO-G (Stage \Rst) across all metrics. This confirms that although DAO-P can generate structures, its generative performance is compromised, which motivates training DAO-G as a separate, specialized generator.
    \item Stage \Rst+\Rnd\, vs. Stage \Rst+\Rnd\,(stable+relax). To verify whether the unstable structures still valuable for training, we remove unstable structures entirely, i.e., only using stable + relaxed data in Stage \Rnd\, pretraining. The results indicate a trade-off between Match Rate and RMSE: performance slightly declines in terms of Match Rate while improving in terms of RMSE. For instance, compared to using the full dataset, the Match Rate decreases from 65.97\% to 65.29\% on MP-20 and from 32.59\% to 32.16\% on MPTS-52, while the RMSE shows a corresponding improvement. This suggests that incorporating unstable (unrelaxed) structures contributes to the generation of more reasonable structures (as reflected in the higher Match Rate), even though it leads to a marginal degradation in RMSE.
    \item Stage \Rst\, vs. Stage \Rst\,(only relax). We conducted an additional experiment where we used an untrained DAO-G model and pretrained it solely on the augmented Stage \Rnd\ dataset. This was done to verify whether the improvements come mainly from DAO-P imputing structures during relaxation. While this approach achieved reasonable results, the performance was consistently lower than that of both our Stage \Rst–pretrained and Stage \Rnd-pretrained DAO-G. For example, compared to DAO-G (Stage~\Rst), the Match Rate dropped from 65.60\% to 58.37\% on MP-20, and from 32.52\% to 23.78\% on MPTS-52. These results indicate that our two-stage pretraining strategy, as well as retaining non-relaxed data in Stage~\Rst, is crucial for our framework's superior performance.
\end{itemize}

\subsubsection{Ablations on Energy Relaxation Threshold}
In Stage \Rnd\, pretraining, we selected only structures with Ehull in the range (0.08,0.5] eV/atom for relaxation. The rationale is as follows:

\begin{itemize}
    \item Structures with Ehull>0.5 eV/atom exhibit poor thermodynamic stability and are far from the ground-state configuration. Relaxing such structures often requires a larger optimization step size, which may adversely affect the optimization of lower-energy structures. Considering both computational efficiency and stability, we excluded structures with Ehull>0.5 eV/atom from relaxation.
    \item The choice of the 0.5 threshold was initially empirical, as it lies near the midpoint of the Ehull distribution in our pretraining dataset ([0, 1.0] eV/atom). 
\end{itemize}

\begin{figure}[t!]
\centering
\includegraphics[scale=0.44]{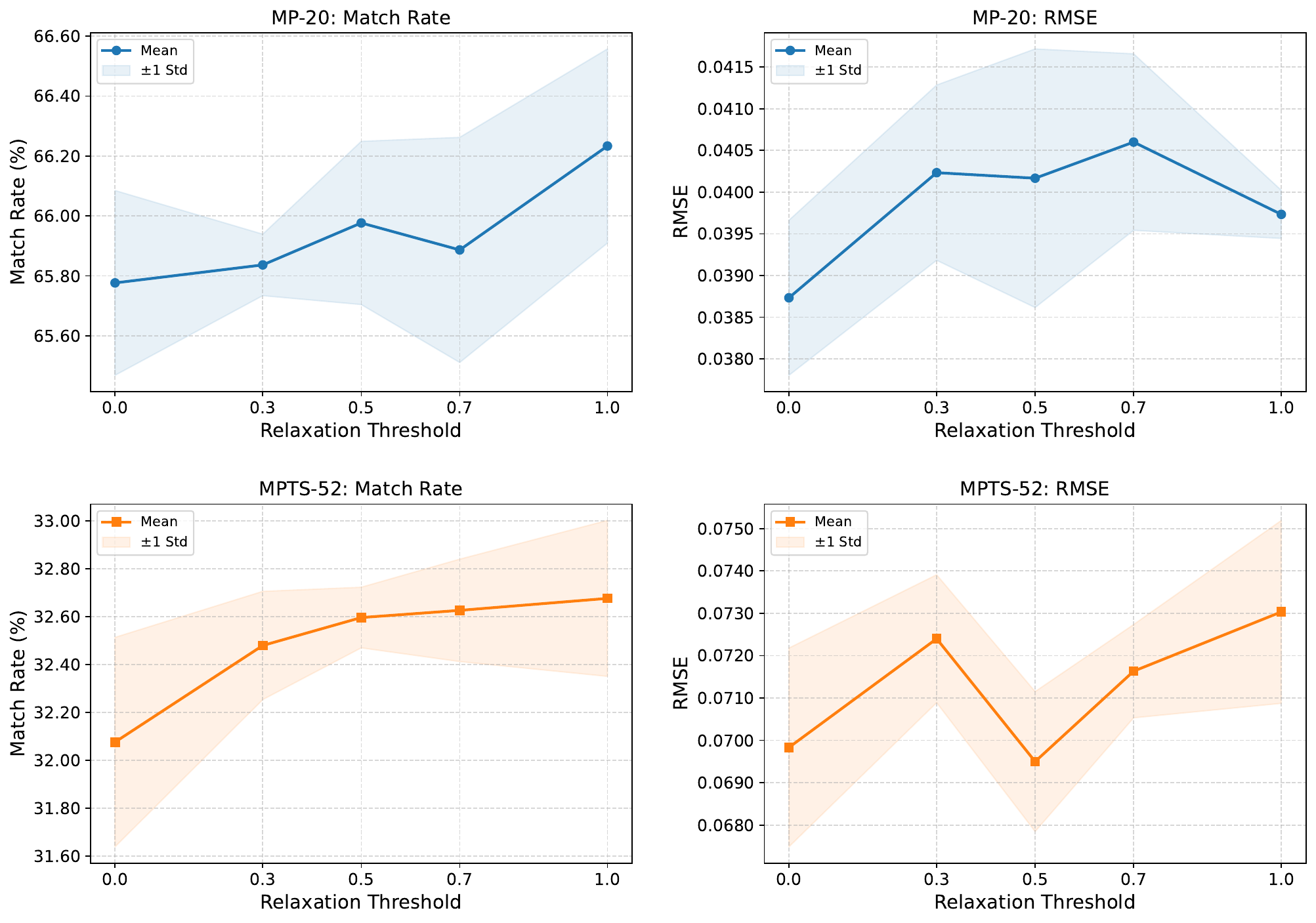}
\caption{The influence of relaxation thresholds. The standard deviations are calculated across three runs.}
\label{fig:abla_relaxation_threshold}
\end{figure}

To further validate this choice, we conducted an ablation study by varying the relaxation threshold among 0.0 (no relaxation), 0.3, 0.5, 0.7, and 1.0 (relax all structures), while keeping the same L-BFGS parameters described in \cref{sec:two_stage_pretraining}. The results (\cref{fig:abla_relaxation_threshold}) show that:

\begin{itemize}
    \item Compared with no relaxation (threshold = 0.0), the threshold of 0.5 yields a higher Match Rate, suggesting that incorporating more relaxed structures in pretraining helps the model generate more physically plausible structures. RMSE is slightly higher, which we attribute to the model’s exposure to more successfully matched structures (as Match Rate is higher), increasing the diversity of  atomic coordinates after relaxation.
    \item For all thresholds greater than zero, the Match Rate increases on both datasets as the threshold is raised. Although a threshold of 1.0 yields the highest Match Rate, it also results in a relatively high RMSE on MPTS-52. Considering the trade-offs between all four metrics across both datasets, we found a threshold of 0.5 to offer the most balanced and reasonable performance.
\end{itemize}

Overall, a threshold of 0.5 strikes the best balance between high Match Rate, low RMSE, and reasonable relaxation cost, as it avoids excessive relaxation of very unstable structures while still improving performance.

Moreover, we have performed a detailed analysis of the structural changes during dataset relaxation. Specifically, we relaxed 429,731 structures with Ehull in the range (0.08, 0.5] eV/atom. After relaxation, we reapplied the same deduplication procedure described in \cref{sec:dataset_dedup}. We found that only 8 relaxed structures matched any entries in the downstream test sets (which contain 17,142 structures in total, i.e., 9,046 in MP-20 and 8,096 in MPTS-52). This proportion is extremely small, suggesting that the impact of dataset relaxation on potential data leakage is negligible.

\subsubsection{20-shot Results}\label{sec:20_shot_results}
We also perform 20-shot sampling for the CSP evaluation and compare DAO-G against several non-pretrained baseline methods. As shown in \cref{tab:20_shot_csp}, DAO-G achieves state-of-the-art performance across all evaluation metrics on both the MP-20 and MPTS-52 datasets. These results demonstrate the effectiveness and superiority of DAO-G under a multiple-sampling setting.

\begin{table*}[t!]
  \centering
  \setlength{\tabcolsep}{10pt}
  \caption{A Comparison of 20-shot sampling Results.}
  \label{tab:20_shot_csp}
  \vspace{0.03in}
  \resizebox{0.94\textwidth}{!}{
    \begin{tabular}{lcccc}
    \toprule
    \multirow{2}[2]{*}{} & \multicolumn{2}{c}{\textbf{MP-20}} & \multicolumn{2}{c}{\textbf{MPTS-52}} \\
             & \textbf{Match Rate (\%) $\uparrow$} & \textbf{RMSE $\downarrow$} & \textbf{Match Rate (\%) $\uparrow$} & \textbf{RMSE $\downarrow$} \\
    \midrule
    \multirow{1}[1]{*}{CDVAE} 
      & 66.95  & 0.1026 & 20.79 & 0.2085 \\
    
    \midrule
    \multirow{1}[1]{*}{DiffCSP}
     & 77.93 &	 0.0492 & 34.02 & 0.1749 \\
    
    \midrule
    
    \multirow{1}[1]{*}{FlowMM}
    &75.81 &\underline{0.0479} &34.05 &0.1813 \\
    \midrule
    \multirow{1}[1]{*}{CrystalFlow}
     &\underline{78.34} &0.0577 &\underline{40.37} &\underline{0.1576}  \\
    \midrule

    \multirow{1}[1]{*}{DAO-G (w/o pretrain)} 
     & 76.49	& 0.0600	&35.44	&0.1447 \\
    
    \addlinespace[2.5pt] 
    \hdashline 
    \addlinespace[2.5pt] 
    
    \multirow{1}[1]{*}{DAO-G} 
     &\textbf{82.68} & \textbf{0.0279} & \textbf{46.78} & \textbf{0.0795} \\
    \bottomrule
    \end{tabular}%
}
\end{table*}

\subsection{Accurate Crystal Property Prediction via the Finetuned DAO-P}\label{sec:exp_property_prediction}

In the main context, DAO-P has functioned as an energy predictor, assisting DAO-G through dataset relaxation and sampling guidance. We now explore whether the pretraining of DAO-P on energy prediction provides a robust foundation for predicting a broader spectrum of material properties. Specifically, we first pretrain DAO-P on the full version of CrysDB and then finetune it on eight representative datasets selected from three widely-used benchmarks: Matbench~\cite{supp:dunn2020_matbench}, JARVIS-3D~\cite{supp:choudhary2020_jarvis3d} and MP~\cite{supp:jain2013_mp_dataset}. To ensure consistency and comparability, we adopt the experimental settings described in Matformer~\cite{supp:yan2022matformer} for JARVIS-3D and MP, and utilize the default settings for the Matbench datasets. For comparison, we choose two categories of SOTA methods on these benchmarks: models without pretraining~\cite{supp:schutt2017schnet,supp:xie2018cgcnn,supp:chen2019megnet,supp:choudhary2021alignn,supp:yan2022matformer,supp:lin2023potnet}, and models with pretraining~\cite{supp:das2023crysgnn,supp:magar2022crystaltwins,supp:yu2023mmpt}. We quantify the accuracy using the MAE metric.

\begin{figure}[t!]
\centering
\includegraphics[scale=0.6]{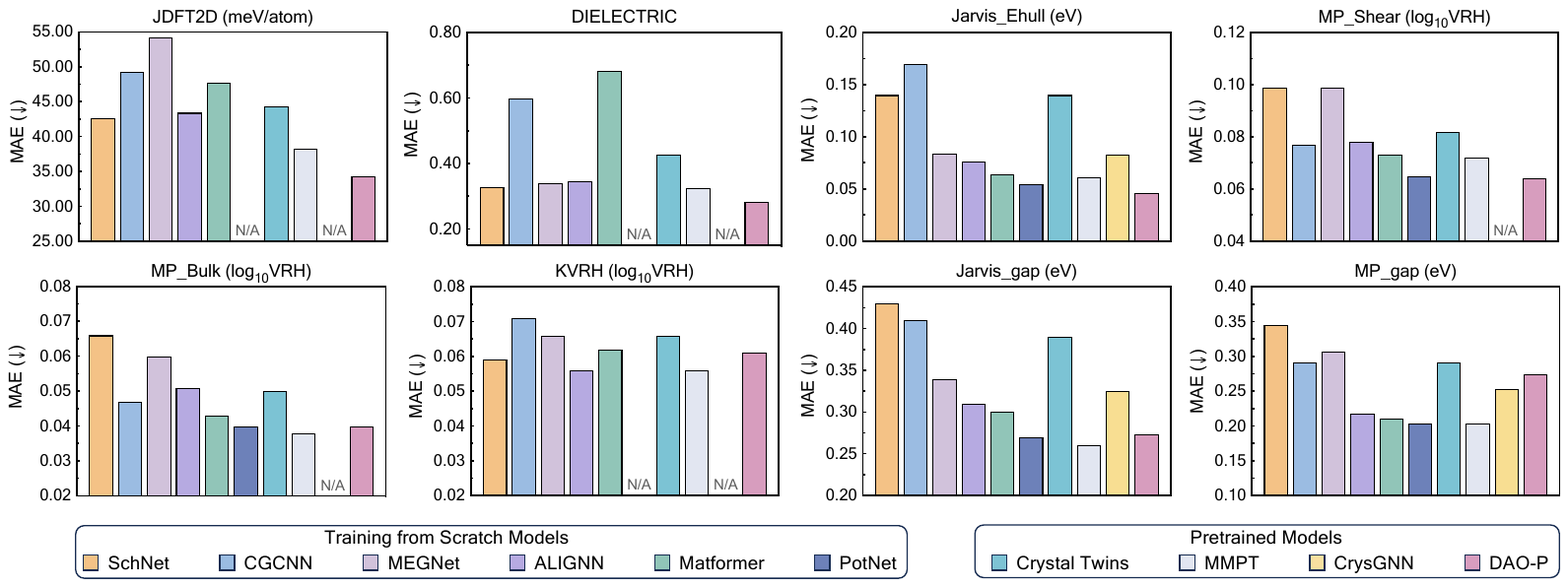}
\caption{The performance of DAO-P for crystal property prediction is evaluated on eight datasets. The compared baselines include models both with and without pretraining. The results for PotNet~\cite{supp:lin2023potnet} and CrysGNN~\cite{supp:das2023crysgnn} are obtained from their original publications, while the results for the remaining baseline models are sourced from the MMPT paper~\cite{supp:yu2023mmpt}. For baselines where the corresponding experiments were not conducted in the original paper, the results are denoted as N/A. The raw results are recorded in \cref{tab:DAO-P_comprison}.}
\label{fig:propery_prediction}
\end{figure}

The results are presented in \cref{fig:propery_prediction}. Notably, DAO-P outperforms previous approaches on half of the evaluated datasets and achieves competitive results on the remaining datasets, with the exception of MP\_gap. Across these datasets, DAO-P consistently ranks among the top three performers, demonstrating its robustness and generalization capability. Particularly striking is the significant performance improvement on Jarvis\_Ehull, where DAO-P achieves a 16.3\% increase over the second-best method. Furthermore, the effectiveness of DAO-P in few-shot learning scenarios is demonstrated by its outstanding performance on the JDFT2D dataset, which contains only 636 instances.
In conclusion, DAO-P is not limited to energy prediction; it can be adapted to predict a wide range of material properties with appropriate finetuning datasets. Moreover, the principle of energy guidance can be extended to other properties. For instance, DAO-P could be utilized to bias the generation process toward structures with enhanced Shear Modulus.

\begin{table}[t!]
\caption{Experimental results (MAE $\downarrow$) of crystal property prediction on eight datasets.}
\label{tab:DAO-P_comprison}
\centering
\renewcommand{\arraystretch}{1.5}
\resizebox{0.98\textwidth}{!}{
\begin{tabular}{ccccccccc}
\toprule
\multirow{2}[2]{*}{} & \multicolumn{3}{c}{\textbf{MatBench}} & \multicolumn{2}{c}{\textbf{JARVIS-3D}} & \multicolumn{3}{c}{\textbf{MP}}  \\ \cmidrule(lr){2-4}
\cmidrule(lr){5-6} \cmidrule(lr){7-9} 
 & \makecell{JDFT2D \\ (meV/atom)} & DIELECTRIC & \makecell{KVRH \\ ($\log_{10}\text{VRH}$)} & \makecell{Jarvis\_gap \\ (eV)} & \makecell{Jarvis\_Ehull \\ (eV)} & \makecell{Mp\_Shear \\ ($\log_{10}\text{VRH}$)} & \makecell{Mp\_Bulk \\ ($\log_{10}\text{VRH}$)} & \makecell{MP\_gap \\ (eV)}  \\ \midrule
SchNet & 42.663 & 0.327 & \underline{0.059} & 0.430 & 0.140 & 0.099 & 0.066 & 0.345  \\
CGCNN & 49.244 & 0.598 & 0.071 & 0.410 & 0.170 & 0.077 & 0.047 & 0.292  \\
MEGNet & 54.171 & 0.339 & 0.066 & 0.340 & 0.084 & 0.099 & 0.060 & 0.307  \\
ALIGNN & 43.424 & 0.344 & \textbf{0.056} & 0.310 & 0.076 & 0.078 & 0.051 & 0.218  \\
Matformer & 47.696 & 0.681 & 0.062 & 0.300 & 0.064 & 0.073 & 0.043 & \underline{0.211}  \\
PotNet & -  & - & - & \underline{0.270} & \underline{0.055} & \underline{0.065} & 0.040 & \textbf{0.204} \\ \hline
Crystal Twins & 44.353 & 0.427 & 0.066 & 0.390 & 0.140 & 0.082 & 0.050 & 0.291  \\
MMPT & \underline{38.213} & \underline{0.324} & \textbf{0.056} & \textbf{0.260} & 0.061 & 0.072 & \textbf{0.038} & \textbf{0.204}  \\
CrysGNN & -  & - & - & 0.325 & 0.083 & - & - & 0.253 \\ \hline
DAO-P  & \textbf{34.280} & \textbf{0.283} & 0.061 &0.273  & \textbf{0.046} &  \textbf{0.064} & \underline{0.040} & 0.275 \\ \bottomrule 
\end{tabular}
}
\end{table}

In addition, we further investigated the impact of pretraining by applying our same pretraining dataset to two of the train-from-scratch baselines: SchNet~\cite{supp:schutt2017schnet} and MEGNet~\cite{supp:chen2019megnet}. It should be noted that the original results of these two models were reproduced from MMPT~\cite{supp:yu2023mmpt}. However, since MMPT is not open-sourced, we are unable to directly access its code for pretraining. As stated in the MMPT appendix, the results on the three Matbench datasets were obtained from its website, while other results were reproduced from Matformer. Although Matformer provides model parameters, it does not release the training code for these two models. As a result, different hyperparameters were used across datasets, and the lack of accessible code posed challenges for reproduction. The primary aim of this experiment is to validate the performance improvement attributable to pretraining. Therefore, it is paramount to maintain identical code implementations both before and after pretraining. For this, we used public versions from PyG and DeepChem to ensure differences are not due to code or configuration. We also scaled their parameters (SchNet: 25.8M, MEGNet: 25.2M) to match our DAO-P model (25.2M) closely.

As a result, this part of the evaluation will differ from \cref{tab:DAO-P_comprison}, and we make this clarification to avoid confusion. The updated results are shown in \cref{tab:prop_baseline_pretrain} and the corresponding training hyperparameters are presented in \cref{tab:prop_baseline_hypers}. Across eight downstream property prediction datasets, both SchNet and MEGNet generally exhibit lower MAE after pretraining (with the exception of SchNet on MP\_Bulk, which shows a slight increase). However, even after pretraining on the same dataset, both models still underperform DAO-P, which highlights the strong effectiveness of our DAO-P model. 

\begin{table}[t!]
\caption{The downstream property prediction results (MAE$\downarrow$) of pretrained baselines.}
\label{tab:prop_baseline_pretrain}
\centering
\renewcommand{\arraystretch}{1.5}
\resizebox{0.96\textwidth}{!}{
    \begin{tabular}{ccccccccc}
        \toprule
\multirow{2}[2]{*}{} & \multicolumn{3}{c}{\textbf{MatBench}} & \multicolumn{2}{c}{\textbf{JARVIS-3D}} & \multicolumn{3}{c}{\textbf{MP}}  \\ \cmidrule(lr){2-4}
\cmidrule(lr){5-6} \cmidrule(lr){7-9} 
 & \makecell{JDFT2D \\ (meV/atom)} & DIELECTRIC & \makecell{KVRH \\ ($\log_{10}\text{VRH}$)} & \makecell{Jarvis\_gap \\ (eV)} & \makecell{Jarvis\_Ehull \\ (eV)} & \makecell{Mp\_Shear \\ ($\log_{10}\text{VRH}$)} & \makecell{Mp\_Bulk \\ ($\log_{10}\text{VRH}$)} & \makecell{MP\_gap \\ (eV)}  \\ \midrule
        SchNet & 65.912 &0.417 &0.104 &0.635 &0.205 &0.106 &0.069 & 0.412  \\
        SchNet (Pretrain) & 63.503 & 0.402 & 0.095 & 0.444 & 0.177 & 0.091 & 0.072 & 0.379  \\ \midrule
        MEGNet & 67.143 & 0.360 & 0.081 & 0.429 & 0.159 & 0.105 & 0.073 & 0.351 \\
        MEGNet (Pretrain) & 54.713 & 0.349 & 0.073 & 0.403 & 0.144 & 0.101 & 0.068 & 0.338 \\ \midrule
        Our & \textbf{34.280} & \textbf{0.283} & \textbf{0.061} & \textbf{0.273} & \textbf{0.046} & \textbf{0.064} & \textbf{0.040} & \textbf{0.275} \\
        \bottomrule 
    \end{tabular}
}
\end{table}

\begin{table}[t!]
\caption{The hyperparameters of SchNet and MEGNet.}
\label{tab:prop_baseline_hypers}
\centering
\renewcommand{\arraystretch}{1.5}
\resizebox{0.96\textwidth}{!}{
    \begin{tabular}{ccccccccc}
        \toprule
\multirow{2}[2]{*}{} & \multicolumn{3}{c}{\textbf{MatBench}} & \multicolumn{2}{c}{\textbf{JARVIS-3D}} & \multicolumn{3}{c}{\textbf{MP}}  \\ \cmidrule(lr){2-4}
\cmidrule(lr){5-6} \cmidrule(lr){7-9} 
 & \makecell{JDFT2D \\ (meV/atom)} & DIELECTRIC & \makecell{KVRH \\ ($\log_{10}\text{VRH}$)} & \makecell{Jarvis\_gap \\ (eV)} & \makecell{Jarvis\_Ehull \\ (eV)} & \makecell{Mp\_Shear \\ ($\log_{10}\text{VRH}$)} & \makecell{Mp\_Bulk \\ ($\log_{10}\text{VRH}$)} & \makecell{MP\_gap \\ (eV)}  \\ \midrule
        SchNet & \makecell{lr=0.0001 \\ decay=0.0001} &\makecell{lr=0.0005 \\ decay=0.0005} &\makecell{lr=0.001 \\ decay=0.0001} &\makecell{lr=0.001 \\ decay=0.001} &\makecell{lr=0.0005 \\ decay=0.0001} &\makecell{lr=0.001 \\ decay=0.0001} &\makecell{lr=0.0003 \\ decay=0.0} & \makecell{lr=0.0005 \\ decay=0.0001}  \\ \midrule
        SchNet (Pretrain) & \makecell{lr=0.0005 \\ decay=0.00005} & \makecell{lr=0.0001 \\ decay=0.0001} & \makecell{lr=0.0001 \\ decay=0.001} & \makecell{lr=0.0001 \\ decay=0.0002} & \makecell{lr=0.0005 \\ decay=0.0001} & \makecell{lr=0.0001 \\ decay=0.00005} & \makecell{lr=0.001 \\ decay=0.00005} & \makecell{lr=0.001 \\ decay=0.0001}  \\ \midrule
        MEGNet & \makecell{lr=0.0001 \\ decay=0.00005} & \makecell{lr=0.001 \\ decay=0.0001} & \makecell{lr=0.001 \\ decay=0.0001} & \makecell{lr=0.001 \\ decay=0.0001} & \makecell{lr=0.0005 \\ decay=0.0001} & \makecell{lr=0.0005 \\ decay=0.0001} & \makecell{lr=0.001 \\ decay=0.0001} & \makecell{lr=0.0005 \\ decay=0.0005} \\ \midrule
        MEGNet (Pretrain) & \makecell{lr=0.001 \\ decay=0.0001} & \makecell{lr=0.001 \\ decay=0.0001} & \makecell{lr=0.0001 \\ decay=0.00005} & \makecell{lr=0.001 \\ decay=0.001} & \makecell{lr=0.0005 \\ decay=0.0001} & \makecell{lr=0.0005 \\ decay=0.0001} & \makecell{lr=0.0005 \\ decay=0.0001} & \makecell{lr=0.001 \\ decay=0.0005} \\ \bottomrule 
    \end{tabular}
}
\end{table}

\subsection{Additional Results on Superconductor Validation}\label{sec:supercon_app}
In this section, we provide further experimental details regarding the validation on superconductors, including implementation specifics (e.g., hyperparameter settings) and corresponding DFT results.

\subsubsection{Joint Superconducting Property Prediction and Structure Generation}
In \cref{sec:supercon3d_aug}, we train DAO-P with and without augmented data, respectively. For each setting, we conduct a grid search over learning rates (2e-4, 3e-4, and 4e-4) and weight decay values (1e-5, 2e-5, 3e-5, and 5e-5), and then select the respective best model for comparison in \cref{tab:fold_results}. As expected, the augmentation with generated structures significantly improves the performance of DAO-P, resulting in the best performance on all five folds.

\begin{table}[t!]
\centering
\caption{The best configuration of each $T_c$ predictor and the corresponding MAE (logK). ``aug.'' denotes the model is augmented by generated structures.}
\label{tab:fold_results}
\renewcommand{\arraystretch}{1.5}
\resizebox{0.99\textwidth}{!}{
\begin{tabular}{lcccccccc}
\toprule
\textbf{Model} & \textbf{LR} & \textbf{Decay} & \textbf{Fold 1} & \textbf{Fold 2} & \textbf{Fold 3} & \textbf{Fold 4} & \textbf{Fold 5} & \textbf{Average} \\
\midrule
DAO-P (w/o aug.) & 2e-4 & 5e-5 &0.838 ± 0.016 &0.755 ± 0.021 &0.694 ± 0.003 &0.763 ± 0.017 &0.758 ± 0.012 &0.761 ± 0.011\\
DAO-P (aug.) & 4e-4 & 5e-5 &\textbf{0.823 ± 0.009} &\textbf{0.695 ± 0.016} &\textbf{0.636 ± 0.006} &\textbf{0.718 ± 0.021} &\textbf{0.699 ± 0.007} &\textbf{0.714 ± 0.004}\\
\bottomrule
\end{tabular}
}
\end{table}

\begin{table}[t!]
\centering
\caption{$T_c$ prediction results on three real-world superconductors.}
\label{tab:tc_prediction}
\begin{tabular}{lccc}
\toprule
\textbf{Formula} & \textbf{Experimental $T_c$ (K)} & \textbf{Predicted $T_c$ (K)} & \textbf{Absolute Error (K)} \\
\midrule
$\text{Cr}_6\text{Os}_2$~\cite{supp:flukiger1974electronically} & 3.99 & 1.97 &2.02\\
$\text{Zr}_{16}\text{Pd}_8\text{O}_4$~\cite{supp:watanabe2023_sc2} & 2.73 & 2.99 &0.26\\
$\text{Zr}_{16}\text{Rh}_8\text{O}_4$~\cite{supp:watanabe2023_sc2} & 3.73 & 3.77 &0.04\\
\bottomrule
\end{tabular}
\end{table}

Additionally, we extend our experiments to three real-world superconductors, utilizing the previously finetuned DAO-G and DAO-P (aug.) models. We first employ the ensemble DAO-P models to predict $T_c$ for them and the results on \cref{tab:tc_prediction} indicate the predicted $T_c$ values are consistent with the experimentally determined values. Then DAO-G is used to generate structures for the three superconductors, given only their formulas, with 20-shot sampling. Further, we select the sample with the lowest RMSE from the generated 20 samples for each superconductor, and visualize them in the second row of \cref{fig:supercon}d. Impressively, all three generated structures show excellent agreement with the ground truth structures. These findings suggest that models finetuned on SuperCon exhibits strong generalization to newly discovered superconductors.

\begin{table}[t!]
\centering
\caption{The configurations of DFT calculation.}
\label{tab:dft_configuration}
\resizebox{0.94\textwidth}{!}{
\begin{tabular}{lccccccc}
\toprule
\multicolumn{4}{c}{\textbf{system}} & \multicolumn{1}{c}{\textbf{electrons}} & \multicolumn{2}{c}{\textbf{cell}} &\multicolumn{1}{c}{\textbf{ions}}  \\ \cmidrule(lr){1-4}
\cmidrule(lr){5-5} \cmidrule(lr){6-7} \cmidrule(lr){8-8} 
ecutwfc & occupations & smearing & degauss & conv\_thr & press\_conv\_thr & cell\_dynamics & ion\_maxstep \\
\midrule
40.0 & smearing & gaussian & 0.01d0 & 1.0d-6 & 0.5 & bfgs & 50 (default) \\
\bottomrule
\end{tabular}
}
\end{table}

\begin{table}[t!]
\centering
\caption{The results of 20 runs of DFT calculations. ``N/A'' denotes the failed match.}
\resizebox{0.96\textwidth}{!}{
\begin{tabular}{|c|c|c|c|c|c|c|c|c|c|c|}
\hline
Run-id & 1 & 2 & 3 & 4 & 5 & 6 & 7 & 8 & 9 & 10 \\ \hline
RMSE   & 0.0414 & 0.2716 & N/A & 0.2961 & 0.0418 & 0.0410 & 0.2062 & 0.0414 & N/A & 0.0412 \\ \hline
Run-id & 11 & 12 & 13 & 14 & 15 & 16 & 17 & 18 & 19 & 20 \\ \hline
RMSE & 0.3375 & N/A & N/A & 0.0416 & N/A & \textbf{0.0410} & 0.0416 & 0.3437 & 0.0412 & 0.1377\\ \hline
\end{tabular}
}
\label{tab:su1_QE_20}
\end{table}

\begin{table}[t!]
\centering
\caption{The results of 20 runs of DAO-G generation.}
\resizebox{0.96\textwidth}{!}{
\begin{tabular}{|c|c|c|c|c|c|c|c|c|c|c|}
\hline
Run-id & 1 & 2 & 3 & 4 & 5 & 6 & 7 & 8 & 9 & 10 \\ \hline
RMSE & 0.0017 & 0.0018 & 0.0015 & 0.0017 & 0.0016 & 0.0017 & 0.0016 & 0.0016 & 0.0018 & 0.0015 \\ \hline
Run-id & 11 & 12 & 13 & 14 & 15 & 16 & 17 & 18 & 19 & 20 \\ \hline
RMSE & 0.0015 & 0.0015 & 0.0019 & 0.0016 & 0.0016 & \textbf{0.0012} & 0.0020 & 0.0016 & 0.0014 & 0.0016 \\ \hline
\end{tabular}
}
\label{tab:su1_DAO-G_20}
\end{table}

\begin{table}[t!]
\centering
\caption{A comparison between DFT and our DAO-G on the generation of  $\text{Cr}_6\text{Os}_2$~\cite{supp:flukiger1974electronically}.}
\label{tab:comparison_dft}
\renewcommand{\arraystretch}{1.5}
\resizebox{0.94\textwidth}{!}{
\begin{tabular}{lcccc}
\toprule
\textbf{Method} & \textbf{Initialized Structure} & \textbf{RMSE} & \textbf{Iterations} & \textbf{Running Time} \\
\midrule
DFT~\cite{supp:giannozzi2017qe} & \makecell{perturbation on the experiment structure \\ ($10\%$ noise for abc, 10\text{\textdegree} noise for $\alpha \beta \gamma$, 0.1 noise for $\rmF$)} & 0.0410 & \makecell{38 scf-steps \\+ 37 bfgs-steps} & 138.74m \\ \hline
DAO-G & absolute noises sampled from $\gU(0,1)$ and $\gN(\vzero,\rmI)$ & 0.0012 &1000 & 1.50m \\
\bottomrule
\end{tabular}
}
\end{table}

\subsubsection{DFT Results for Three Real-World Superconductors}

\begin{figure}[htbp!]
\centering
\includegraphics[scale=0.65]{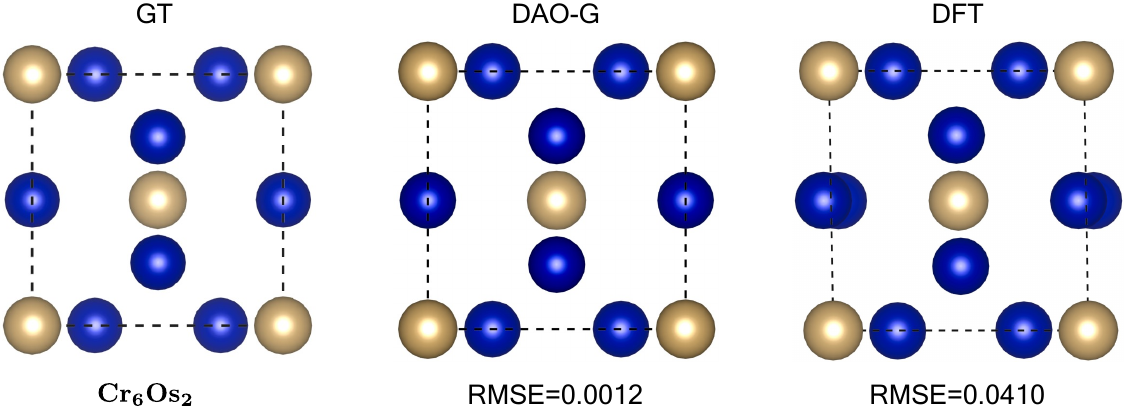}
\caption{An illustration of the structure relaxed by DFT and the structure generated by DAO-G, for the superconductor $\text{Cr}_6\text{Os}_2$~\cite{supp:flukiger1974electronically}. GT is the ground truth structure.}
\label{fig:comparison_dft}
\end{figure}

We additionally employ the Quantum-Espresso software (QE~\cite{supp:giannozzi2017qe}), a prevailing DFT tool, to calculate the relaxed structures for the three real-world superconductors. The configurations of QE calculation are presented in \cref{tab:dft_configuration}. To ensure a fair comparison, the input structures are randomly initialized, consistent with our diffusion-based generation process. However, all three generation attempts are unsuccessful. We hypothesize that this failure stems from the structural complexity of these superconductors. Therefore, to reduce the computational burden, we perturb the ground-truth structures slightly before employing them as starting points for QE calculations. Specifically, we apply 10\% lattice lengths perturbation, 10-degree angles perturbation, and 0.1 fractional coordinates deviation. For $\text{Cr}_6\text{Os}_2$~\cite{supp:flukiger1974electronically}, across 20 runs, QE calculations results in five failed matches, whereas DAO-G achieves success in all runs. As shown in \cref{tab:su1_QE_20,tab:su1_DAO-G_20}, the best result generated by DAO-G is 0.0012, which is significantly superior to the best QE calculation result of 0.0410.
The visualizations of the corresponding best results are depicted in \cref{fig:comparison_dft}. Clear discrepancies exist between the relaxed structure and the ground truth, while the structure generated by DAO-G shows well alignment. 
Interestingly, although DAO-G has seen unstable $\text{Cr}_6\text{Os}_2$ structures during pretraining (no data leakage), it does not bias toward reproducing these unstable configurations at generation time. Instead, DAO-G generates the stable superconducting structure, with a DFT-calculated Ehull of 0.02918, closely matching the experimental value (0.02916). This behavior indicates that DAO-G is not merely memorizing training examples, but has learned a meaningful distribution over stable crystal structures conditioned on composition, even when unstable polymorphs of the same formula are present in the training data.

Regarding efficiency, DAO-G is approximately 92 times faster than DFT calculations. DAO-G generates a structure in 1.5 minutes, while QE requires approximately 2 hours and 18 minutes. The comparison is presented in \cref{tab:comparison_dft}. For the two more complex superconductors ($\text{Zr}_{16}\text{Pd}_8\text{O}_4$~\cite{supp:watanabe2023_sc2} and $\text{Zr}_{16}\text{Rh}_8\text{O}_4$~\cite{supp:watanabe2023_sc2}), even slight perturbations similar to those applied to $\text{Cr}_6\text{Os}_2$ results in computational exceptions. Subsequently, we further decreased the perturbations to 5\% for lattice parameters, 3 degrees for angles, and a standard deviation of 0.05 for fractional coordinates. However, each iteration still requires over 8 hours of computation time, and the Self-Consistent Field (SCF) process fails to converge within 8 days, exceeding practical time constraints.

\section{Visualization}
\subsection{Visualization of Generated Polymorphs}\label{sec:Polymorph_more_vis}
\begin{figure}[t!]
\centering
\includegraphics[scale=1.]{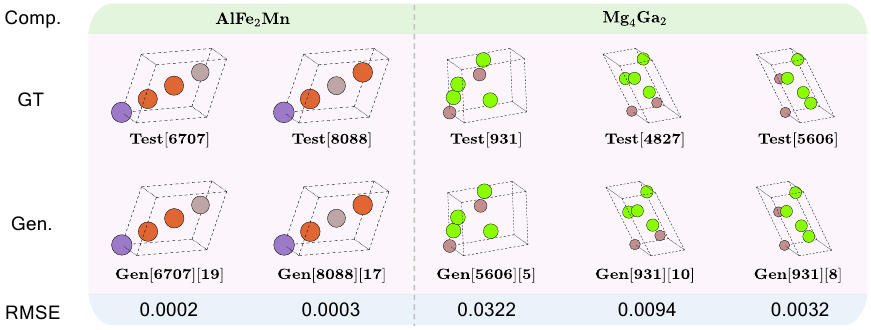}
\caption{The visualization of the polymorphs (with 2 and 3 conformations) generated by DAO-G, with the corresponding ground-truth structures. Comp. = Composition, GT = Ground Truth, Gen. = Generation.}
\label{fig:poly_stats}
\end{figure}

In \cref{fig:poly_stats}, we also provide visualizations for cases involving 2- and 3-polymorphs. The results demonstrate that our model successfully generates diverse structural configurations for each case, highlighting its strong capability in modeling complex polymorphic distributions.

\subsection{Visualization of the Diffusion Process}
\begin{figure}[t!]
\centering
\includegraphics[scale=0.48]{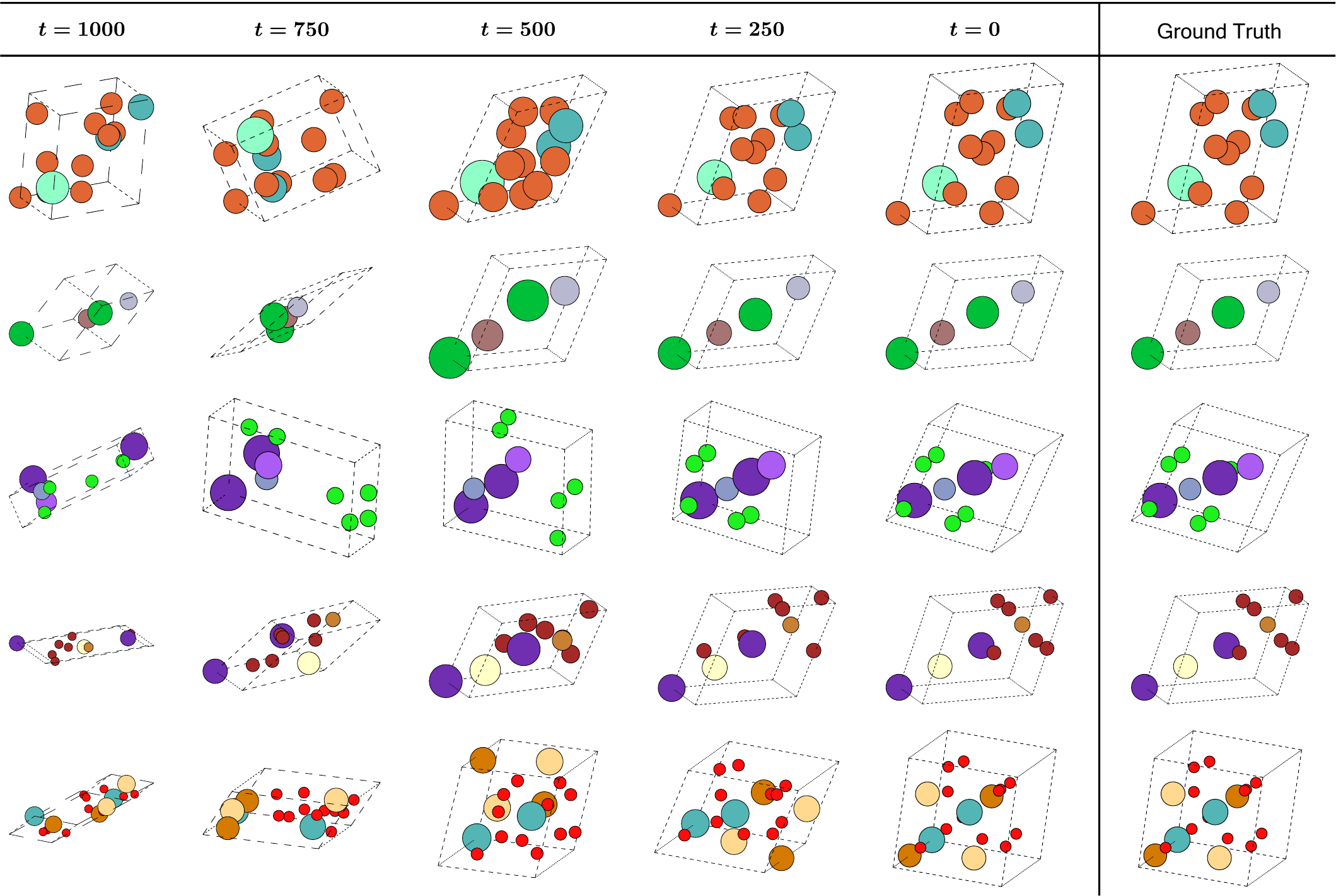}
\caption{Visualizations of the generated structures by DAO-G throughout the diffusion process. We show representative structures at timesteps 1000, 750, 500, 250, and 0. The structures at timestep 0 represent the final generated samples, which are well-aligned with the corresponding ground truth structures. To enhance visual clarity and facilitate comparison, a common atom within each group (represented by a row) has been translated to the origin.}
\label{fig:trajectory}
\end{figure}

To better understand how the structures evolve during the generation process, we select several examples from MP-20 dataset and visualize the perturbed structures at different timesteps in \cref{fig:trajectory}.

\section{Comparison with DiffCSP and MatterGen}
Our method, DAO, shares certain similarities with both DiffCSP~\cite{supp:jiao2024diffcsp} and MatterGen~\cite{supp:zeni2023mattergen}. To further clarify our contributions, this section elaborates on the key distinctions between DAO and each of these approaches.

\subsection{Comparative Analysis with DiffCSP}
Although our method is closely linked to DiffCSP, important differences exist. Below we discuss both the commonalities and distinctions.
\begin{itemize}
    \item \textbf{Relevance.} Both our DAO-G and DiffCSP target the CSP task and employ a joint diffusion process to generate both lattice and fractional coordinates. 
    \item \textbf{Differences.} Despite these shared ideas, our framework substantially differs from DiffCSP in three aspects:
    \begin{enumerate}
        \item \textbf{Training paradigm.} DiffCSP is trained entirely from scratch, while our framework follows a pretrain–finetune paradigm. We conduct a two-stage pretraining on a large-scale crystal dataset (CrysDB, ~910k structures with both structures and energies) and then finetune on the DiffCSP benchmark datasets, leading to stronger performance.
        \item \textbf{Backbone architecture.} DiffCSP adopts a GNN-based backbone (similar to EGNN~\cite{supp:satorras2021egnn}), whereas our framework introduces a Transformer-based backbone (Crysformer), which provides stronger expressivity and generalization ability, especially under large-scale pretraining.
        \item \textbf{Energy modeling.} DiffCSP briefly discusses energy optimization (Appendix G.3), where the energy at intermediate steps is supervised directly using the ground-truth energy at $t=0$, i.e.,
        $\mathcal{L}_\text{DiffCSP}=\|f_\phi(\mathcal{M}_t,t)-\mathcal{E}_0(\mathcal{M}_0))\|_2^2$. This approach has known limitations (presented in Section 4.5), such as inaccurate prediction or numerical instability. In contrast, we propose a new exponential-based energy loss: $\mathcal{L}_\text{DAO}=\|e^{-f_\phi(\mathcal{M}_t,t)}-e^{-\mathcal{E}_0(\mathcal{M}_0)})\|_2^2$, which we have justified theoretically as the optimal formulation (see \cref{sec:loss_design}).
    \end{enumerate}
\end{itemize}

\subsection{Comparative Analysis with MatterGen}\label{sec:dao_vs_mattergen}
MatterGen is a well-known work in crystal generation and is relevant to our study. However, we would like to emphasize that MatterGen’s pretraining primarily targets \emph{de novo}  generation (DNG), where both composition and structure are generated jointly. In contrast, DAO is designed for Crystal Structure Prediction (CSP), where the composition is given and the goal is to generate the corresponding structure. While MatterGen can be adapted for CSP, this is only a minor aspect explored in their work and is not the main focus.

On the other hand, Our DAO-P model and the classifier-free guidance (CFG) approach in MatterGen both provide guidance but differ fundamentally in their nature and purpose:
\begin{enumerate}
    \item \textbf{Model type:} DAO-P is an explicit energy predictor, a neural network trained to estimate the energy of intermediate structures in the reverse process of diffusion. In contrast, CFG, despite its name including classifier, is not an explicit predictor; it is a training strategy that does not require a separate model to make predictions.
    \item \textbf{Underlying principle:} The difference in working mechanisms is closely tied to the tasks each method addresses. Our work focuses on the Crystal Structure Prediction (CSP) task, where the goal is to generate stable crystal structures with low energy. As a result, DAO-P is designed to predict the energy and computes its gradient with respect to the structure, using this information to guide generation toward lower-energy configurations. In MatterGen, however, the goal is conditional crystal generation (e.g., conditioned on chemistry and symmetry), and CFG embeds the provided conditions into the generative model to guide the output toward satisfying those conditions. 
    \item \textbf{Implementation details:} DAO-P takes a crystal structure as input and outputs its predicted energy. CFG, on the other hand, encodes the condition via a lightweight adapter module, concatenates this encoding with the structure’s representation, and feeds the combined input into the generative model. During training, conditions are randomly dropped with a certain probability, but there is still only one generative model plus the adapter module—no separate predictor is involved.    
\end{enumerate}

\putbib[supp_reference]
\label{LastAppendixPage}
\end{bibunit}

\end{document}